\RequirePackage{fix-cm} 
\documentclass[a4paper, twoside, reqno, 12pt]{amsart}
\usepackage{fixltx2e}   

\usepackage{etex}

\usepackage[latin1]{inputenc}
\usepackage[T1]{fontenc}

\usepackage{esint}
\usepackage{dsfont}
\usepackage{xspace}
\usepackage{amsgen}
\usepackage{amsthm}
\usepackage{amssymb}
\usepackage{amsmath}
\usepackage{wasysym}
\usepackage{upgreek}
\usepackage{amsfonts}
\usepackage{stmaryrd}
\usepackage{mathtools}

\usepackage[mathcal, mathscr]{euscript}

\usepackage{mathrsfs}
\DeclareMathAlphabet{\mathscrbf}{OMS}{mdugm}{b}{n}

\usepackage[lined, boxed, linesnumbered, english]{algorithm2e}

\usepackage{multicol}
\usepackage{multirow}

\usepackage{a4wide}

\usepackage[dvipsnames, table]{xcolor}
\definecolor{bckg}{RGB}{20.8, 20.8, 20.8}
\definecolor{oneblue}{rgb}{0.0, 0.0, 0.85}
\definecolor{Lightblue}{RGB}{214, 214, 214}
\definecolor{bluepigment}{rgb}{0.2, 0.2, 0.6}
\definecolor{charcoal}{rgb}{0.21, 0.27, 0.31}
\definecolor{denimblue}{rgb}{0.08, 0.38, 0.74}
\definecolor{Lightgray}{rgb}{0.89, 0.89, 0.89}
\definecolor{darkgrey}{rgb}{0.273, 0.281, 0.30}
\definecolor{darkelectricblue}{rgb}{0.33, 0.41, 0.47}

\usepackage[sort&compress, comma, square, numbers]{natbib}

\usepackage{psfrag}
\usepackage{graphicx}
\usepackage{subfigure}
\usepackage{morefloats}
\usepackage{indentfirst}

\usepackage{acronym}
\usepackage{microtype}
\usepackage[labelsep=period,%
            labelfont={bf,sf,color=bluepigment},%
            justification=raggedright]{caption}

\usepackage[perpage, symbol]{footmisc}

\usepackage[usenames, pdf]{pstricks}
\usepackage{epsfig}
\usepackage{pst-grad} 
\usepackage{pst-plot} 

\usepackage{url}
\usepackage[colorlinks,
           urlcolor=oneblue,
           linkcolor=denimblue,
           citecolor=NavyBlue,
           bookmarksopen=false,
           pdfpagemode=UseNone,
           pagebackref]{hyperref}

\usepackage[explicit]{titlesec}

\titleformat{\section}[block]
  {\color{NavyBlue}\Large\sffamily\bfseries}
  {}
  {0.0em}
  {\colorbox{bckg!5}{\strut\parbox{\dimexpr\linewidth-2\fboxsep\relax}{\thesection. #1}}}
  [\vspace*{0.33em}]

\titleformat{name=\section,numberless}[block]
  {\color{NavyBlue}\Large\sffamily\bfseries}
  {}
  {0.0em}
  {\colorbox{bckg!5}{\strut\parbox{\dimexpr\linewidth-2\fboxsep\relax}{#1}}}
  [\vspace*{0.33em}]

\titleformat{\subsection}
  {\color{NavyBlue}\large\sffamily\bfseries}
  {}
  {0.0em}
  {\colorbox{bckg!5}{\parbox{\dimexpr\linewidth-2\fboxsep\relax}{\thesubsection. #1}}}
  [\vspace*{0.33em}]

\titleformat{name=\subsection,numberless}
  {\color{NavyBlue}\Large\sffamily\bfseries}
  {}
  {0em}
  {\colorbox{bckg!5}{\parbox{\dimexpr\linewidth-2\fboxsep\relax}{#1}}}
  [\vspace*{0.33em}]

\titleformat{\subsubsection}
  {\color{bluepigment}\sffamily\normalsize\bfseries}
  {\thesubsubsection}
  {0.5em}
  {#1}
  [\vspace*{0.33em}]

\titleformat{\paragraph}[runin]
  {\color{bluepigment}\sffamily\small\bfseries}
  {}
  {0em}
  {#1}

\titlespacing{\section}{0.0em}{1.5em plus 2pt minus 2pt}%
{1.0em plus 2pt minus 2pt}[0em]
\titlespacing{\subsection}{0.5em}{1.5em plus 2pt minus 2pt}%
{1.0em}[0em]
\titlespacing{\subsubsection}{0.5em}{1.5em plus 2pt minus 2pt}%
{1.0em plus 2pt minus 2pt}[0em]

\usepackage{titletoc}

\contentsmargin{0.5em}
\newlength{\tocsep} 
\titlecontents{section}[\tocsep]
  {\addvspace{10pt}\bfseries\sffamily}
  {\contentslabel[\thecontentslabel]{\tocsep}}
  {}
  {\ \titlerule*[0.75pc]{.}\ \thecontentspage}
  []
\titlecontents{subsection}[\tocsep]
  {\addvspace{8pt}\sffamily}
  {\contentslabel[\thecontentslabel]{\tocsep}}
  {}
  {\ \titlerule*[0.5pc]{.}\ \thecontentspage}
  []
\titlecontents*{subsubsection}[\tocsep]
  {\addvspace{2pt}\footnotesize\sffamily}
  {}
  {}
  {\ \titlerule*[0.35pc]{.}\ \thecontentspage}
  [\\*]

\makeatletter
\def\@setauthors{%
  \begingroup
  \def\thanks{\protect\thanks@warning}%
  \trivlist
  \centering\footnotesize \@topsep30\p@\relax
  \advance\@topsep by -\baselineskip
  \item\relax
  \author@andify\authors
  \def\\{\protect\linebreak}%
  \textsc{\normalsize\textcolor{darkelectricblue}{\authors}}%
  \ifx\@empty\contribs
  \else
    ,\penalty-3 \space \@setcontribs
    \@closetoccontribs
  \fi
  \endtrivlist
  \endgroup
}
\def\@settitle{\begin{center}%
  \baselineskip14\p@\relax
    \bfseries
    \textsc{\Large\textcolor{charcoal}{\@title}}
  \end{center}%
}
\makeatother

\usepackage{enumitem}
\setlist[description]{%
  topsep=30pt,               
  itemsep=5pt,               
  font={\bfseries\sffamily\color{NavyBlue}}, 
}

\usepackage{fancyhdr}
\usepackage{lastpage}

\newcommand*\Title{\textcolor{bluepigment}{Advanced reduced--order models}}
\newcommand*\Authors{\textcolor{bluepigment}{S.~Gasparin, J.~Berger, D.~Dutykh \& N.~Mendes}}
\newcommand*{\plogo}{\textcolor{gray}{{\texttt{arXiv.org} / \textsc{hal}}}} 

\numberwithin{equation}{section}

\newcommand{\eg}{\emph{e.g.}\xspace}
\newcommand{\etc}{\emph{etc.}\xspace}

\newcommand{\unit}[1]{\, \mathsf{#1} \,}

\newcommand{\Eu}{\textsc{Euler}}
\newcommand{\svd}{SVD}
\newcommand{\deim}{DEIM}

\newcommand{\egal}{\ =\ }
\newcommand{\plus}{\ +\ }
\newcommand{\moins}{\ -\ }

\renewcommand{\O}{\mathcal{O}\, }
\newcommand*{\Ox}{\Omega_{\, x}}

\renewcommand{\b}{\mathrm{b \,}}

\newcommand{\Mat}{\mathrm{Mat}\,}

\newcommand{\BivL}{\mathrm{Bi}_{\,v,\,\mathrm{L}}}
\newcommand{\BivR}{\mathrm{Bi}_{\,v,\,\mathrm{R}}}
\newcommand{\cm}{c_{\,m}}
\newcommand{\cms}{c_{\,m}^{\,\star}}
\newcommand{\dm}{d_{\,m}}
\newcommand{\dmref}{d_{\,m}^{\,0}}
\newcommand{\dms}{d_{\,m}^{\,\star}}

\newcommand{\glL}{g_{\,l,\,\mathrm{L}}}
\newcommand{\glR}{g_{\,l,\,\mathrm{R}}}
\newcommand{\glsL}{g_{\,l, \,\mathrm{L}}^{\,\star}}
\newcommand{\glsR}{g_{\,l,\, \mathrm{R}}^{\,\star}}
\newcommand{\hv}{h_{\,v}}
\newcommand{\hvL}{h_{\,v,\,\mathrm{L}}}
\newcommand{\hvR}{h_{\,v,\,\mathrm{R}}}
\newcommand{\kl}{k_{\,l}}
\newcommand{\kv}{k_{\,v}}
\newcommand{\Pc}{P_{\,c}}
\newcommand{\Ps}{P_{\,s}}
\newcommand{\Pv}{P_{\,v}}
\newcommand{\Pvi}{P_{\,v}^{\,i}}
\newcommand{\PvL}{P_{\,v,\,\mathrm{L}}}
\newcommand{\PvR}{P_{\,v, \,\mathrm{R}}}
\newcommand{\Rv}{R_{\,v}}
\newcommand{\tref}{t^{\,0}}
\newcommand{\uL}{u_{\,\mathrm{L}}}
\newcommand{\uR}{u_{\,\mathrm{R}}}
\newcommand{\rholv}{\rho_{\,l+v}}
\newcommand{\xs}{x^{\,\star}}
\newcommand{\ts}{t^{\,\star}}
\newcommand{\M}{\mathcal{M}}
\newcommand{\A}{\mathcal{A}}
\newcommand{\R}{\mathds{R}}

\newcommand{\scal}{\boldsymbol{\cdot}}
\newcommand*\pd[2]{\frac{\partial #1}{\partial #2}}
\renewcommand{\div}{\grad\scal}
\newcommand{\grad}{\boldsymbol{\nabla}}
\newcommand{\eqdef}{\mathop{\ \stackrel{\mathrm{def}}{:=}\ }}

\begin{document}

\title[\Title]{Advanced reduced-order models for moisture diffusion in porous media}

\author[S.~Gasparin]{Suelen Gasparin$^*$}
\address{\textbf{S.~Gasparin:} LAMA, UMR 5127 CNRS, Universit\'e Savoie Mont Blanc, Campus Scientifique, F-73376 Le Bourget-du-Lac Cedex, France and Thermal Systems Laboratory, Mechanical Engineering Graduate Program, Pontifical Catholic University of Paran\'a, Rua Imaculada Concei\c{c}\~{a}o, 1155, CEP: 80215-901, Curitiba -- Paran\'a, Brazil}
\email{suelengasparin@hotmail.com}
\urladdr{https://www.researchgate.net/profile/Suelen\_Gasparin/}
\thanks{$^*$ Corresponding author}

\author[J.~Berger]{Julien Berger}
\address{\textbf{J.~Berger:} LOCIE, UMR 5271 CNRS, Universit\'e Savoie Mont Blanc, Campus Scientifique, F-73376 Le Bourget-du-Lac Cedex, France}
\email{Berger.Julien@univ-smb.fr}
\urladdr{https://www.researchgate.net/profile/Julien\_Berger3/}

\author[D.~Dutykh]{Denys Dutykh}
\address{\textbf{D.~Dutykh:} Univ. Grenoble Alpes, Univ. Savoie Mont Blanc, CNRS, LAMA, 73000 Chamb\'ery, France and LAMA, UMR 5127 CNRS, Universit\'e Savoie Mont Blanc, Campus Scientifique, F-73376 Le Bourget-du-Lac Cedex, France}
\email{Denys.Dutykh@univ-smb.fr}
\urladdr{http://www.denys-dutykh.com/}

\author[N.~Mendes]{Nathan Mendes}
\address{\textbf{N.~Mendes:} Thermal Systems Laboratory, Mechanical Engineering Graduate Program, Pontifical Catholic University of Paran\'a, Rua Imaculada Concei\c{c}\~{a}o, 1155, CEP: 80215-901, Curitiba -- Paran\'a, Brazil}
\email{Nathan.Mendes@pucpr.edu.br}
\urladdr{https://www.researchgate.net/profile/Nathan\_Mendes/}


\begin{titlepage}
\thispagestyle{empty} 
\noindent
{\Large Suelen \textsc{Gasparin}}\\
{\it\textcolor{gray}{Pontifical Catholic University of Paran\'a, Brazil}}\\
{\it\textcolor{gray}{LAMA--CNRS, Universit\'e Savoie Mont Blanc, France}}
\\[0.02\textheight]
{\Large Julien \textsc{Berger}}\\
{\it\textcolor{gray}{LOCIE--CNRS, Universit\'e Savoie Mont Blanc, France}}
\\[0.02\textheight]
{\Large Denys \textsc{Dutykh}}\\
{\it\textcolor{gray}{LAMA--CNRS, Universit\'e Savoie Mont Blanc, France}}
\\[0.02\textheight]
{\Large Nathan \textsc{Mendes}}\\
{\it\textcolor{gray}{Pontifical Catholic University of Paran\'a, Brazil}}
\\[0.10\textheight]

\colorbox{Lightblue}{
  \parbox[t]{1.0\textwidth}{
    \centering\huge\sc
    \vspace*{0.7cm}
    
    \textcolor{bluepigment}{Advanced reduced--order models for moisture diffusion in porous media}

    \vspace*{0.7cm}
  }
}

\vfill 

\raggedleft     
{\large \plogo} 
\end{titlepage}


\newpage
\thispagestyle{empty} 
\par\vspace*{\fill}   
\begin{flushright} 
{\textcolor{denimblue}{\textsc{Last modified:}} \today}
\end{flushright}


\newpage
\maketitle
\thispagestyle{empty}


\begin{abstract}

It is of great concern to produce numerically efficient methods for moisture diffusion through porous media, capable of accurately calculate moisture distribution with a reduced computational effort. In this way, model reduction methods are promising approaches to bring a solution to this issue since they do not degrade the physical model and provide a significant reduction of computational cost. Therefore, this article explores in details the capabilities of two model-reduction techniques - the Spectral Reduced-Order Model (Spectral--ROM) and the Proper Generalised Decomposition (PGD) --- to numerically solve moisture diffusive transfer through porous materials. Both approaches are applied to three different problems to provide clear examples of the construction and use of these reduced-order models. The methodology of both approaches is explained extensively so that the article can be used as a numerical benchmark by anyone interested in building a reduced-order model for diffusion problems in porous materials. Linear and non-linear unsteady behaviors of unidimensional moisture diffusion are investigated. The last case focuses on solving a parametric problem in which the solution depends on space, time and the diffusivity properties. Results have highlighted that both methods provide accurate solutions and enable to reduce significantly the order of the model around ten times lower than the large original model. It also allows an efficient computation of the physical phenomena with an error lower than $10^{\,-2}$ when compared to a reference solution.


\bigskip
\noindent \textbf{\keywordsname:} reduced-order modelling;  moisture diffusion; numerical methods; spectral methods; Proper Generalised Decomposition (PGD) \\


\end{abstract}


\newpage
\tableofcontents
\thispagestyle{empty}


\newpage
\section{Introduction}

Mathematical models representing the physical phenomena of heat and moisture transfer in porous media have been developed since the works of \textsc{Philip} and \textsc{De Vries} \cite{Philip1957} and \textsc{Luikov} \cite{Luikov1966}. According to \textsc{Luikov} \cite[Chapter 6]{Luikov1966}, the following system of differential equations represents the physical phenomenon of heat and mass transfer through capillary porous materials:
\begin{subequations}\label{eq:Luikov}
\begin{align}
  \pd{U}{t} & \egal \div \bigl(\, a_{\,m} \, \grad U \plus \delta \, a_{\,m} \grad T \,\bigr) \,, \\
  c_{\,b} \, \rho_{\,0} \pd{T}{t} & \egal \div \bigl(\, \lambda \grad T \,\bigr) \plus r_{\,12} \div \biggl(\, a_{\,m1} \, \rho_{\,0} \bigl(\, \grad U \plus \delta_{\,1} \grad T \,\bigr) \,\biggr)\,,
\end{align}
\end{subequations}
where $U$ is the specific mass content in the porous body, corresponding to the ratio of the mass of the water (in the vapor and the liquid phase)  to the mass of dry-basis body. The other quantities are $T\,$, the temperature, $a_{\,m}\,$, the mass transfer coefficient for vapor (denoted with the subscript $1$) and liquid inside the body, $\delta\,$, the thermal-gradient coefficient, $\rho_{\,0}\,$, the specific mass of the dry body, $c_{\,b}\,$, the specific heat of the body and, $r_{\,12}\,$, the latent heat of vaporization.

Particularly, in the context of building components, ones aims at estimating accurately the moisture transfer since it is related to the occupants' health, to the durability of building porous elements and to the building energy consumption and demand. However, the elaboration of numerical methods for the mathematical problem~\eqref{eq:Luikov} faces several issues. First, the physical phenomena of moisture and heat transfer do not evolve on the same time scale. The characteristic time may scales, for many materials, from minutes for the heat diffusion and to a month for the moisture diffusion. Moreover, the time horizon of the studies is of the order of $1\ \mathsf{year}$ when evaluating the building energy consumption. The characteristic times of the physical phenomena have differences of several orders of magnitude. Another complexity of the problem arises from the material properties, which strongly vary with moisture and temperature, with space when considering multi-layers porous walls and sometimes with times when looking at the durability of the materials. Therefore, these complexities challenge the elaboration of numerical models.

Several works have been presented in the literature to propose numerical methods for some mathematical model~\eqref{eq:Luikov}. A list of the main tools can be obtained in \cite{Mendes2017, Woloszyn2008}. Most of the hygrothermal simulation tools are built using numerical approaches and discrete representations of the continuous equations by means of standard and incremental techniques (Finite-Difference, Finite-Volume and Finite-Element methods) to compute the solution. Due to stability conditions, most of the approaches are based on implicit schemes as described for instance in \cite{Mendes2005, Janssen2007}. Therefore, these approaches do not circumvent the complexities of the mathematical problem. These schemes require the solution of large systems of equations (an order of $10^{\,6}$ for three-dimensional problems). Moreover, when considering nonlinear building material properties, sub-iterations at each time step are induced, increasing significantly the computational cost as mentioned for examples in \cite{DosSantos2006, Dalgliesh2005}. It is important to mention that the computational resources available in a computer are not increasing anymore \cite{Waldrop2016}. Thus, it is worth investigations to develop numerical models based on an optimal usage of the available computational resources.

Therefore, the challenge relies on proposing efficient numerical models, capable of computing accurately the solution with a reduced computational effort. These models are of great concern particularly for investigations requiring important computations, as for instance sensitivity analysis or parameter estimation problems. In this way, model reduction techniques appear as a very interesting alternative to substantially reduce the number of operations, saving computational resources (CPU time and memory) with no loss of accuracy of the solution. It is important to note that the fidelity of the mathematical model is conserved.  In recent years, reduced-order modeling techniques have proven to be powerful tools, providing accurate predictions while dramatically reducing computational time, for a wide range of applications, covering from fluid mechanics, heat transfer, structural dynamics and other fields.

Therefore, the scope of this work is limited to the application of two advanced \emph{a priori} reduction-order model techniques, the PGD \cite{Berger2015} and the Spectral--ROM \cite{Gasparin2018} techniques, which have been successfully implemented for coupled heat and moisture diffusion problems. The comparison is carried out for moisture transfer phenomena providing an important primary evaluation for anyone intending to build a reduced-order model for diffusion problem. 
Several features of the methods are analyzed in terms of both model order reduction and accuracy of the computed solution. The investigation is carried for three case studies: (\textit{i}) linear transfer; (\textit{ii}) parametric problems, aiming at computing a model whose solution depends on the material properties and (\textit{iii}) nonlinear transfer with moisture dependent material properties.

Finally, the manuscript is organized as follows: first, Section~\ref{sec:physical_phenom} presents the description of the physical phenomena; then Section~\ref{sec:methods_decription} gives an explanation of the PGD and Spectral reduced-order model techniques; and, further sections present the three different case studies: (\textit{i}) linear transfer in Section~\ref{sec:AN1}; (\textit{ii}) parametric problem whose depends on time, space and on the diffusion parameter in Section~\ref{sec:AN2}, and (\textit{iii}) nonlinear transfer in Section~\ref{sec:AN3}.


\section{Moisture transfer in porous materials}
\label{sec:physical_phenom}

The physical problem involves unidimensional moisture diffusion through a porous material defined by the spatial domain $\Ox \,=\, [\, 0\,,  L \,] $. The moisture transfer only occurs according to the liquid and vapor diffusion. The physical problem can be formulated as \cite{Janssen2014, AitOumeziane2014, Roels2003}: 
\begin{align}\label{eq:moisture_equation_1D}
  & \pd{\rholv}{t} \egal \pd{}{x} \; \left( \, \kl \, \pd{\Pc}{x} \plus \kv \, \pd{\Pv}{x} \, \right) \,,
\end{align}
where $\rholv$ is the volumetric moisture content of the material and $\kv$ and $\kl$ are the vapor and liquid permeabilities.

Eq.~\eqref{eq:moisture_equation_1D} can be written using the vapor pressure $\Pv$ as the driving potential. For this, we consider the physical relation, known as the \textsc{Kelvin} equation, between $\Pv$ and $\Pc\,$:
\begin{align*}
  \Pc & \egal \Rv \, T \, \rho_{\,l}\, \ln\left(\frac{\Pv}{\Ps(T)}\right)\,, \\
  \pd{\Pc}{\Pv} & \egal \frac{R_{\,v} \, T\, \rho_{\,l}}{\Pv} \,.
\end{align*}
Thus, we have:
\begin{align*}
  \pd{\Pc}{x} \egal \pd{\Pc}{\Pv} \cdot \pd{\Pv}{x} \plus \pd{\Pc}{T} \cdot \pd{T}{x} \,.
\end{align*}
As we consider the mass transfer under isothermal conditions, the second term vanishes and we obtain: 
\begin{align*}
  \pd{\Pc}{x} \egal  \frac{\Rv \, T\, \rho_{\,l}}{\Pv} \cdot \pd{\Pv}{x} \,.
\end{align*}

In addition, we have:
\begin{align*}
  \pd{\rholv}{t} \egal \pd{\rholv}{\phi} \cdot \pd{\phi}{\Pv} \cdot \pd{\Pv}{t} \plus \pd{\rholv}{T} \cdot \pd{T}{t} \ \simeq \ \pd{\rholv}{\phi} \cdot \pd{\phi}{\Pv} \cdot \pd{\Pv}{t} \,.
\end{align*}
Considering the relation $\rholv \egal f\, (\phi) \egal f\, (\, \Pv \,,T \,)\,$, obtained from material properties and from the relation between the vapour pressure $\Pv$ and the relative humidity $\phi\,$, we get: 
\begin{align*}
  \pd{\rholv}{t} \egal f^{\,\prime}(\phi) \; \frac{1}{\Ps} \; \pd{\Pv}{t} \,.
\end{align*}
Eq.~\eqref{eq:moisture_equation_1D} can be therefore rewritten as:
\begin{align}\label{eq:moisture_equation_1D_v2}
  & f^{\,\prime}(\phi) \, \frac{1}{\Ps} \cdot \pd{\Pv}{t} \egal \pd{}{x} \; \biggl[ \, \biggl( \, \kl \, \frac{\Rv \, T\, \rho_{\,l}}{\Pv} \plus \kv \, \biggr) \, \pd{\Pv}{x} \, \biggr] \,.
\end{align}
The material properties $f\,$, $\kl$ and $\kv$ depend on the vapor pressure $\Pv\,$. Therefore, we denote $\dm \eqdef \kl \, \dfrac{\Rv \, T\, \rho_{\,l}}{\Pv} \plus \kv $ as the global moisture transport coefficient and $\cm \eqdef f^{\,\prime}(\phi) \; \dfrac{1}{\Ps}$ the moisture storage coefficient.

At the material bounding surfaces, \textsc{Robin}-type boundary conditions are considered:
\begin{subequations}\label{eq:bc}
\begin{align}
  \dm \, \pd{\Pv}{x} &\egal \hvL \cdot \left( \, \Pv \moins \PvL \, \right) \moins \glL \, , && x \egal 0 \,, \\[3mm]
  -\ \dm \, \pd{\Pv}{x} &\egal \hvR \cdot \left( \, \Pv \moins \PvR \, \right) \moins \glR \, ,&& x \egal L \,,
\end{align}
\end{subequations}
where $\PvL$ and $\PvR$ are the vapor pressure of the ambient air, $\glL$ and $\glR$ are the liquid flow (driving rain) at the two bounding surfaces.
The initial condition is consider with a uniform vapor pressure distribution:
\begin{align*}
  \Pv &\egal \Pvi \,, && t\egal0 \,.
\end{align*}

It is important to obtain a unitless formulation of governing equations while performing mathematical and numerical analysis of given practical problems, due to a certain number of reasons already discussed in \cite{Gasparin2017}. Therefore, we define the following dimensionless parameters:
\begin{align*}
& u \egal \frac{\Pv}{\Pvi} \,,
&& \uR \egal \frac{\PvR}{\Pvi} \,,
&& \uL \egal \frac{\PvL}{\Pvi} \,,
&& \xs \egal \frac{x}{L} \,, \\[3pt]
& \ts \egal \frac{t}{\tref} \,,
&& \cms \egal \frac{\cm \cdot L^{\,2}}{\dmref \cdot \tref} \,,
&& \dms \egal \frac{\dm}{\dmref} \,,
&& \BivL \egal \frac{\hvL \cdot L}{\dmref}  \,,\\[3pt]
& \BivR \egal \frac{\hvR \cdot L}{\dmref} \,,
&& \glsL \egal \frac{\glL \cdot L}{\dmref \cdot \Pvi} \,,
&& \glsR \egal \frac{\glR \cdot L}{\dmref \cdot \Pvi}\,.
\end{align*}
In this way, the dimensionless governing equations are then written as:
\begin{subequations}\label{eq:moisture_dimensionlesspb_1D}
\begin{align}
 \cms \pd{u}{\ts} &\egal \pd{}{\xs}\; \left( \, \dms \; \pd{u}{\xs} \, \right) \,,
& \ts & \ > \ 0\,, \;&  \xs & \ \in \ \big[ \, 0, \, 1 \, \big] \,, \\[3pt]
 \dms \, \pd{u}{\xs} &\egal \BivL \cdot \left( \, u \moins \uL \, \right) \moins \glsL \,,
& \ts & \ > \ 0\,, \,&  \xs & \egal 0 \,, \\[3pt]
\moins \dms \, \pd{u}{\xs} &\egal \BivR \cdot \left( \, u \moins \uR \, \right) \moins \glsR \,,
& \ts & \ > \ 0\,, \,&   \xs & \egal 1 \,, \\[3pt]
 u &\egal 1 \,,
& \ts & \egal 0\,, \,&  \xs & \ \in \ \big[ \, 0, \, 1 \, \big] \,.
\end{align}
\end{subequations}
Finally, this is the problem of interest considered in this work. The procedure of the methods used for the problem solution is described in the next section.


\section{Methodology}
\label{sec:methods_decription}

For the sake of simplicity, without losing the generality, the methods are first explained considering $\dms$ and $\cms$ as constants, noting $\nu \eqdef \dfrac{\dms}{\cms} \,$, and we drop $\star$ for the sake of notation compactness. Thus, considering a linear diffusion equation:
\begin{align}\label{eq:heat1d}
  \pd{u}{t} \egal \nu \, \pd{}{x}  \biggl( \, \pd{u}{x} \, \biggr) \,,
\end{align}
for $x \, \in \, \big[\, 0, \, 1 \, \big] $. The boundary conditions are:
\begin{subequations}\label{eq:bc_heateq}
\begin{align}
  \pd{u}{x}  &\egal \BivL\, \Bigl(u \moins \uL\, (\, t \,)\Bigr) \,, &  x & \egal 0 \,, \\
  \moins \pd{u}{x} &\egal \BivR\, \Bigl(u \moins \uR\, (\, t \,)\Bigl)\,, &  x & \egal 1 \,.
\end{align}
\end{subequations}
Using a standard discretization method, such as \textsc{Euler} or \textsc{Crank}--\textsc{Nicolson}, to compute the solution of Eq.~\eqref{eq:heat1d} yields in computing a solution $u \, (\,x\,,t\,)$ for each point of the discretised spatial and time domains. The following is adopted: $N_{\,x}$ and $N_{\,t}\,$, which stand for the number of elements according to the discretization of the space and time domains. Thus, the order of the so-called large original model is $p \egal N_{\,x} \cdot N_{\,t}\,$.

Model reduction aims at decreasing the degrees of freedom present in a numerical model. It aims at approximating the solution by a lower order model $N \, \ll \, p$ without reducing drastically the fidelity of the physical model. One of the features is to significantly decrease the computational resources space (CPU time and memory).

The scope of this work is limited to the two \emph{a priori} reduction-order model techniques, the PGD \cite{Berger2015} and the Spectral--ROM \cite{Gasparin2018} techniques, which have been successfully implemented in building physics to reduce computational cost while maintaining high fidelity solutions. Both techniques assume separated tensorial representation of the solution by a finite sum of function products. The Spectral--ROM fixes a set of spatial basis functions to be the \textsc{Chebyshev} polynomials and then, a system of ordinary differential equations is built to compute the temporal coefficients of the solution using the \textsc{Tau}--\textsc{Galerkin} method, while the PGD aims at computing directly the basis of functions by minimizing the residual. Further details on both methodologies are given in the next sections.


\subsection{Reduced Spectral method}

Spectral methods are successfully applied in studies of wave propagation,  meteorology, computational fluid dynamics, quantum mechanics and several other fields \cite{Canuto2006}.  Some works on the transport phenomena can be found in literature involving diffusive \cite{Guo2012, Wang2016}, convective \cite{Chen2016, RamReddy2015} and radiative \cite{Li2008a, Chen2015b, Ma2014} heat transfer. Spectral techniques applied in these works are varied, adopted according to the geometry, boundary conditions, field of application, and other necessities. In recent works, researchers have implemented spectral methods for solving heat and moisture transfer in food engineering \cite{Pasban2017} and on fluid flows \cite{Motsa2015}. According to the authors's knowledge, there are no results in the literature regarding the application of spectral methods for solving diffusive moisture transfer in building physics applications other than \cite{Gasparin2018}.

Spectral methods consider a global representation of the solution, which means the derivative at a certain spatial point depends on the solution of the entire domain and not only on its neighbors \cite{Boyd2000}. Besides, Spectral methods consider a sum of polynomials that suit for the whole domain, almost like an analytical solution, providing a high approximation of the solution. Therefore, as its error decreases exponentially, it is possible to have the same accuracy of other methods but with a lower number of \textsc{Galerkin} modes, which makes this method memory minimizing, allowing to store and operate a lower number of degrees of freedom \cite{Trefethen1996}.


\subsubsection{Application of the Spectral--ROM}

The idea of the Spectral method is to assume that the unknown $u\, (\,x\,, t\,)$ from Eq.~\eqref{eq:heat1d} can be approximatively represented as a finite sum \cite{Mendes2017}:
\begin{equation}\label{eq:series_ap}
  u\, \, (\,x\,,t\,) \ \approx\ u_{\, n}\, \, (\,x\,,t\,) \egal \sum_{i\, =\, 0}^n \, a_{\,i}\, (\,t\,)\, \phi_{\,i}\, (\,x\,)\,.
\end{equation}
Here, $\{\phi_{\,i}\, (\,x\,)\}_{\,i\, =\, 0}^{\,n}$ is a set of basis functions that remains constant in time, $\{a_{\,i}\, (\,t\,)\}_{\,i\, =\, 0}^{\,n}$ are the corresponding time-dependent Spectral coefficients, and $n$ represents the number of degrees of freedom of the solution. Eq.~\eqref{eq:series_ap} can be seen as a series truncation after $N \, = \, n\, +\, 1$ modes \cite{Gasparin2018}. The \textsc{Chebyshev} polynomials are chosen as the basis functions as they are optimal in $\mathcal{L}_{\, \infty}$ approximation norm \cite{Gautschi2004}. Therefore, we have:
\begin{align*}
  \phi_{\,i}\, (\,x\,)\ \equiv\ T_{\,i}\, (\,x\,)\,.
\end{align*}
For more details on \textsc{Chebyshev} polynomials the readers may refer to \cite{Peyret2002, Boyd2000}.

As we have chosen the basis functions, we can compute the derivatives:
\begin{subequations}\label{eq:derivatives}
\begin{align}
  \pd{u_{\,n}}{x} &\egal \sum_{i\, =\, 0}^n \, a_{\,i}\,(\,t\,)\, \pd{T_{\,i}}{x}\,(\,x\,)\egal \sum_{i\, =\, 0}^n \tilde{a}_{\,i}\,(\,t\,)\, T_{\,i}\,(\,x\,)\,,\label{eq:derivative1}\\
  \pd{^{\,2} u_{\,n}}{x^{\,2}} &\egal \sum_{i\, =\, 0}^n \, a_{\,i}\,(\,t\,)\, \pd{^{\,2} T_{\,i}}{x^{\,2}}\,(\,x\,)\egal \sum_{i\, =\, 0}^n \Tilde{\Tilde{a}}_{\,i}\,(\,t\,)\, T_{\,i}\,(\,x\,) \,, \label{eq:derivative2}\\
  \pd{u_{\,n}}{t} &\egal \sum_{i\, =\, 0}^n \, \dot{a}_{\,i}\,(\,t\,)\, T_{\,i}\,(\,x\,)\,,\label{eq:derivative3} 
\end{align}
\end{subequations}
where the dot denotes $\dot{a}_{\,i}\, (\,t\,) \eqdef \dfrac{\mathrm{d}\,a\, (\,t\,) }{\mathrm{d}\,t} $ according to \textsc{Newton}'s notation. Note that the derivatives are re-expanded in the same \textsc{Chebyshev} basis function. As a result, coefficients $\{\tilde{a}_{\,i}\, (\,t\,)\}$ and $\{\Tilde{\Tilde{a}}_{\,i}\, (\,t\,)\}$ must be re-expressed in terms of coefficients $\{a_{\,i}\, (\,t\,)\}$. The connection is given explicitly from the recurrence relation of the \textsc{Chebyshev} polynomial derivatives \cite{Peyret2002}.

Using the expression of the derivatives provided by Eqs.~\eqref{eq:derivative2} and \eqref{eq:derivative3}, the residual of the diffusion equation~\eqref{eq:heat1d} is written: 
\begin{align}\label{eq:heat_eq_residual}
  R\,(\,x\,,\,t \,) \egal \sum_{i\, =\, 0}^n \, \Bigl[\, \dot{a}_{\,i}\, (\,t\,) \moins \nu \ \Tilde{\Tilde{a}}_{\,i}\, (\,t\,)\, \Bigr]\, T_{\,i}\, (\,x\,)\,,
\end{align}
which is considered a misfit of the approximate solution. The purpose is to minimize the residual:
\begin{align*}
  \Bigl\Vert\;  R\,(\,x\,,\,t \,)\; \Bigr\Vert \ \longrightarrow\ \min \,,
\end{align*}
which is solved via the \textsc{Tau}--\textsc{Galerkin} method that requires Eq.~\eqref{eq:heat_eq_residual} to be orthogonal to the \textsc{Chebyshev} basis functions $\langle\,R\,,T_{\,i} \,\rangle \,=\, 0$. The scalar product is defined by:
\begin{align*}
  \langle\,f\,,g \,\rangle \egal \int_{-1}^{1}\, \dfrac{f\,(\,x\,)\,g\,(\,x\,)}{\sqrt{1 \moins x^{\,2}}}\; \mathrm{d} x \,.
\end{align*}
Thus, it leads to the following relation between the spectral coefficients: 
\begin{align*}
  \dot{a}_{\,i}\, (\,t\,) \moins \nu \, \Tilde{\Tilde{a}}_{\,i}\, (\,t\,) \egal 0\,, & & i \egal 0, \,1, \,\ldots,\, n-2 \,.
\end{align*}

Finally, after the projection and expansion of the residual, the result is a system of Ordinary Differential Equations (ODE), with $N\, -\, 2$ equations to be solved as a function of time. The two extra coefficients are obtained by substituting the derivative~\eqref{eq:derivative1} into the boundary conditions~\eqref{eq:bc_heateq}. A special attention must be given to the spatial domain, because the \textsc{Chebyshev} Spectral method we use is described between the interval $\big[ -1\,,  1 \, \big]$. Thus, if the dimensionless interval is not in this interval, a change of variables (domain transformation) must be performed for the computational domain. Therefore, two more equations are added to the system:
\begin{subequations}
\begin{align}
  \sum_{i\, =\, 0}^n \, \tilde{a}_{\,i}\, (\,t\,)\, T_{\,i}\, (-1) \moins \BivL \, \sum_{i\, =\, 0}^n \, a_{\,i}\,(\,t\,)\, T_{\,i}\,(-1) \plus \BivL \, \uL\, (\,t\,)  &\egal 0\,, \label{eq:bc1_aproxi} \\
  \moins \sum_{i\, =\, 0}^n \, \tilde{a}_{\,i}\, (\,t\,)\, T_{\,i}\, (\, 1\,) \moins \BivR \, \sum_{i\, =\, 0}^n \, a_{\,i}\, (\,t\,)\, T_{\,i}\, (\, 1\,) \plus \BivR \,\uR\, (\,t\,) &\egal 0\,, \label{eq:bc2_aproxi}
\end{align}
\end{subequations}
with $T_{\,i}\,(-1)\, =\, (-1)^{\,i}$ and $T_{\,i}\,(\, 1\,)\, \equiv\, 1$ (see \cite{Peyret2002}). Eqs.~\eqref{eq:bc1_aproxi} and \eqref{eq:bc2_aproxi} are written in an explicit way, with coefficients $a_{\,n}$ and $a_{\, n-1}$ expressed in terms of the other coefficients.

Therefore, the original partial differential equation~\eqref{eq:heat1d} is reduced to a system of ODEs plus two algebraic expressions. For linear problems, the system of ODEs can be explicitly built. Otherwise, we have a system of Differential-Algebraic Equations. Moreover, the reduced system of ordinary differential equations has the following form:
\begin{align}\label{eq:system_ODE}
  \dot{a}_{\,i}\, (\,t\,) \egal \A \, a_{\,i}\, (\,t\,) \plus \b(\,t\,)\,,  & & i \egal 0, \,1, \,\ldots,\, n-2 \,,
\end{align}
where, $\b(\,t\,) \in \R^{\,N}$ is a vector coming usually from boundary conditions and $\A \in \Mat_{N\times N}(\R )\,$ is a matrix with constant coefficients and with $\O(\, N\,) \simeq 10\,$. The main advantage of a Spectral--ROM is that $N \ll p\,$, where $p$ is the number of degrees of freedom needed to solve problem~\eqref{eq:bc_heateq} by means of conventional methods (finite-differences, finite-elements and finite-volumes).

Initial values of the coefficients $\{a_{\,i}\, (t\, =\, 0)\}$ are calculated by the \textsc{Galerkin} projection of the initial condition \cite{Canuto2006}:
\begin{align}\label{eq:system_ODE_int}
  a_{\,0,\,i}\ \equiv\ a_{\,i}\, (\, 0\,) \egal \dfrac{2}{\pi\, c_{\,i}}\int_{-1}^{\,1}\dfrac{u_{\, 0}\, (\,x\,)\, T_{\,i}\, (\,x\,)}{\sqrt{1 \moins x^{\,2}}}\, \mathrm{d}x\,, & & i \egal 0, \,1, \,\ldots,\, n-2 \,,
\end{align}
where, $u_{\, 0}\, (\,x\,)$, is the dimensionless initial condition. After solving the \emph{reduced} system of ODEs (Eqs.~\eqref{eq:system_ODE} and \eqref{eq:system_ODE_int}), it is possible to compose the solution along with the \textsc{Chebyshev} polynomial.

Thus, by using the Spectral--ROM approach to build the reduced-order model, the time-dependent coefficients $\{a_{\,i}\,(\,t\,) \}$ are computed by solving the following system:
\begin{equation}\label{eq:system_ode}
  \left\{ \begin{array}{rcl}
    \dot{a}_{\,i}\, (\,t\,) & \egal& \A\, a_{\,i}\, (\,t\,) \plus \b\, (\,t\,) \,, \\
    a_{\,i}\, (\, 0\,)& \egal& a_{\,0,\,i} \,,
  \end{array}\right.
\end{equation}
We note that the matrix $\A$ and the vector $\b\,(\,t\,)$ might depend on problem parameters, such as the diffusion coefficient $\nu\,$:
\begin{align*}
  \A \egal \A \, (\, t \,; \nu\,)\,, \ \ \text{and} \ \ \b \egal \b(\, t \,; \nu\,) \,.
\end{align*}

Different approaches can be used to solve the system of ODEs~\eqref{eq:system_ode}, depending on the cases considered. The most straightforward way to solve it is to apply a numerical integration scheme, \eg, an adaptive \textsc{Runge}--\textsc{Kutta} with moderate accuracy. So, with an embedded error control and not so stringent tolerances, it can be done very efficiently. A detailed presentation on how to solve this system can be found in Appendix~\ref{sec:append_solving_ode}.


\subsection{Proper Generalised Decomposition}
\label{sec:PGD}

The Proper Generalised Decomposition (PGD) is also a model reduction method. It originates in the radial space-time separated representation proposed by \textsc{Ladev\`eze} in $1985$ \cite{Ladeveze1985}. In $2006\,$, the separated representations were extended to the multidimensional case by \textsc{Chinesta} and co-workers \cite{Ammar2007}. Interested readers may see \cite{Chinesta2011, Chinesta2013} for additional details on the method as well as \cite{Chinesta2013a} for an introduction. This strategy has been successfully applied and validated for various industrial applications. For instance, the PGD method was applied to quantum mechanics (\textsc{Schr\"odinger} equation) \cite{Ammar2008}, kinetics theory (\textsc{Fokker--Planck} equation non-\textsc{Newtonian} fluids) \cite{Pruliere2010, Ammar2010}, phase separation in heterogeneous mixtures (\textsc{Langer} equation) \cite{Lamari2012}, virtual surgery (forces, vibrations, \etc) \cite{Niroomandi2012}, nonlinear stochastic problems (\textsc{Burgers} equation, 2D nonlinear diffusion problems) \cite{Nouy2007}, multi-scale and multiphysics problems (visco plasticity, damage, \etc) \cite{Neron2010, Bognet2012}, computational fluid dynamics (anisotherm \textsc{Navier}--\textsc{Stokes} problems) \cite{Dumon2011}, and, more recently to, heat and moisture transfer in building materials \cite{Berger2016c}.

The PGD solution to problem Eq.~\eqref{eq:heat1d} is sought as a separated representation of functions of time $t$ and space $x\,$:
\begin{align}\label{eq:aprox_PGD}
  u \, (\,x \,, t \,)\ \simeq\ \sum_{i\,=\, 1}^{M} F^{\,i} \, (\, x \,) \ G^{\,i} \, (\, t \,) \,.
\end{align}
The order of PGD ROM scales with $\O(\,p\,) \egal M \cdot \bigl(\, N_{\,x} \plus N_{\,t} \, \bigr)$.


\subsubsection{Iterative resolution}

Solving problem~\eqref{eq:heat1d} numerically using the PGD method consists in calculating modes $(F^{\,i} \,, G^{\,i})$ iteratively from $i \egal 1$ to $M\,$. The first mode $(F^{\,1} \,, G^{\,1})$ is initialised in order to satisfy the initial and boundary conditions in all zones. At enrichment step $m\, <\, M$, we assume that a former approximation of $u\,(\,x\,,t\,)$ is known and the new couple $F^{\,m+1} \, (\,x \,) \egal R \, (\, x \,)$ and $G^{\,m+1} \, (\, t \,) \egal S \, (\,t \,)$ has to be calculated according to: 
\begin{equation}\label{eq:enrichissement}
  u \, (\,x \,, t \,) \egal \sum_{i\,=\, 1}^{m} \, F^{\,i}\, (\,x\,) \, G^{\,i}\, (\,t\,) \plus R\, (\,x\,) \, S\, (\,t\,) \,.
\end{equation}
Eq.~\eqref{eq:enrichissement} is introduced into Eq.~\eqref{eq:heat1d}. Thanks to the separated representation of the solution $u\,(\,x\,,t\,)$ for dimensions $t$ and $x$ (Eq.~\eqref{eq:aprox_PGD}), we get: 
\begin{equation}\label{eq:enrichissement_equation}
\begin{split} 
  \frac{\mathrm{d} S}{ \mathrm{d} t} \ R \moins \nu \ S \ \frac{\mathrm{d}^{\,2} R}{ \mathrm{d} x} \egal \sum_{i\,=\,1}^{m} \,\frac{\mathrm{d} G^{\,i}}{ \mathrm{d} t} \ F^{\,i} \moins \nu \ G^{\,i} \ \frac{\mathrm{d}^{\,2} F^{\,i}}{ \mathrm{d} x^{\,2}} \plus \mathrm{Res}^{\,m+1} \,,
\end{split}
\end{equation}
where $ \mathrm{Res}^{\,m+1}$ is the residual of Eq.~\eqref{eq:heat1d} because Eq.~\eqref{eq:enrichissement} is an approximation of the solution.


\subsubsection{Computing R(t) and S(x)}

We are at step $m$ searching for the new couple $R$ and $S$ by solving Eq.~\eqref{eq:enrichissement_equation}. To compute them, Eq.~\eqref{eq:enrichissement_equation} will be successively projected on $R$ and $S\,$. For this, we note the scalar product $\langle \, \bullet , \, \bullet \rangle_{\, y}$ in the domain $\Omega_{\,y}\,$, defined by:
\begin{align*}
  \langle \, f \,,\, g \, \rangle_{\, y} \egal \int_{\Omega_{\,y}} f \, g \; \mathrm{d}y \,,
\end{align*}
evaluated numerically using the discrete values of functions $f$ and $g$ and a trapezoidal approximation for the integral. Here, the scalar product is defined for both time and space domains.

Eq.~\eqref{eq:enrichissement_equation} is projected on $R\,$, assuming $\langle \, \mathrm{Res}^{\,m+1}, R \, \rangle_{\, x} \egal 0$ to obtain: 
\begin{equation}\label{eq:equation_R}
  \alpha_{\,1} \ \frac{\mathrm{d} S}{ \mathrm{d} t} \moins \beta_{\,1} \ S \egal \gamma_{\,1} \,,
\end{equation}
with:
\begin{multline*}
  \alpha_{\,1} \egal \langle \, R \,, R \, \rangle_{\, x} \,,\qquad \beta_{\,1} \egal  \biggl\langle \, R \,,  \nu \, \frac{\mathrm{d}^{\,2} R}{ \mathrm{d} x^{\,2}} \, \biggr\rangle_{\, x}  \,, \\ 
  \gamma_{\,1} \egal \displaystyle \sum_{i\,=\,1}^{m} \langle \,   R\,, - F^{\,i} \,  \rangle_{\, x} \ \frac{\mathrm{d} G^{\,i}}{ \mathrm{d} t} +  \biggl\langle R \,,  \nu \, \frac{\mathrm{d}^{\,2} F^{\,i}}{ \mathrm{d} x^{\,2}} \, \biggr\rangle_{\, x} \ G^{\,i} \,.
\end{multline*}
Eq.~\eqref{eq:enrichissement_equation} is now projected on $S\,$, assuming $\langle \, \mathrm{Res}^{\,m+1} \,, S \rangle_{\, t} \egal 0$ and gives: 
\begin{equation}\label{eq:equation_S}
  \alpha_{\,2} \ R \plus \beta_{\,2} \ \frac{\mathrm{d}^{\,2} R}{ \mathrm{d} x^{\,2}} \egal \gamma_{\,2} \,,
\end{equation}
with:
\begin{multline*}
  \alpha_{\,2} \egal \biggl\langle \, S \,, \frac{\mathrm{d} S}{ \mathrm{d} t} \, \biggr\rangle_{\, t} \,, \qquad \beta_{\,2} \egal  \langle \, S \,, S \, \rangle_{\, t} \,, \\
  \gamma_{\,2} \egal \displaystyle \sum_{i\,=\,1}^{m} \, \biggl\langle \, S \,, - \frac{\mathrm{d} G^{\,i}}{ \mathrm{d} t} \, \biggr\rangle_{\, t} \ F^{\,i}  \plus \langle \, S \,, G^{\,i} \,\rangle_{\, t} \ \frac{\mathrm{d}^{\,2} F^{\,i}}{ \mathrm{d} x^{\,2}} \,.
\end{multline*}
After theses projections, to solve Eqs.~\eqref{eq:equation_R} and \eqref{eq:equation_S}, an alternating direction fixed-point algorithm is used. The stopping criterion, assuming the algorithm has converged, is:
\begin{align*}
  \Bigl\Vert\, R^{\,q} \moins R^{\,q-1} \,\Bigr\Vert \ \leqslant \ \eta_{\,1}\ \text{ and }\  \Bigl\Vert\, S^{\,q} \moins S^{\,q-1} \,\Bigr\Vert \ \leqslant \ \eta_{\,1} \,,
\end{align*}
where $q$ is the index of iteration of the fixed-point algorithm and $\eta_{\,1}$ is a tolerance parameter chosen by the user.


\subsubsection{Convergence of the global enrichment}

Functions $R$ and $S$ have just been computed by a fixed-point algorithm. The PGD basis is enriched, noting $F^{\,m+1} \egal R$ and $G^{\,m+1} \egal S$ the new modes. The field of interest $u$ can be written as: 
\begin{equation*}
  u \, (\,x \,, t \,) \egal \sum_{i\,=\,1}^{m}\, F^{\,i}\,(\,x\,) \  G^{\,i}\,(\,t\,) \plus R\,(\,x\,) \ S\,(\,t\,) \egal \sum_{i\,=\,1}^{m+1}\, F^{\,i}(\,x\,) \  G^{\,i}\,(\,t\,) \,.
\end{equation*}
The enrichment of the PGD solution stops when the norm of the residual $\bigl|\bigl|\, \mathrm{Res}^{\,m+1}  \,\bigr|\bigr|$ is assumed negligible with respect to $\eta_{\,2}\,$, another tolerance parameter chosen by the user:
\begin{align*}
  \Bigl\Vert \, \mathrm{Res}^{\,m+1}  \,\Bigl\Vert \egal \Biggl\Vert\, \sum_{i\,=\,1}^{m} \, \frac{\mathrm{d} G^{\,i}}{ \mathrm{d} t} \ F^{\,i} \moins \nu \ G^{\,i} \ \frac{\mathrm{d}^{\,2} F^{\,i}}{ \mathrm{d} x^{\,2}} \,\Biggr\Vert \ \leqslant \ \eta_{\,2} \,.
\end{align*}


\subsection{Comparison of the numerical solution}

To compare and validate the proposed methods, the error between the solution $u^{\, \mathrm{num}}\,(\,x\,,t\,)$, obtained by one of numerical the methods, and the reference solution $u^{\, \mathrm{ref}}\,(\,x\,,t\,)$, is computed as a function of $x$ by the following formulation:
\begin{align*}
  \varepsilon_{\,2}\, (\, x\,)\ &\eqdef \sqrt{\,\frac{1}{N_{\,t}} \, \sum_{j\, =\, 1}^{N_{\,t}} \, \left( \, u_{\, j}^{\, \mathrm{num}}\, (\,x \,, t \,) \moins u_{\, j}^{\mathrm{\, ref}}\, (\,x \,, t \,) \, \right)^{\,2}}\,,
\end{align*}
where $N_{\,t}$ is the number of temporal steps. The global uniform error $\varepsilon_{\, \infty}$ is given by the maximum value of $\varepsilon_{\,2}\, (\, x\,)\,$: 
\begin{align*}
  \varepsilon_{\, \infty}\ &\eqdef \sup_{x \ \in \ \bigl[\, 0 \,,\, L \,\bigr]} \, \varepsilon_{\,2}\, (\, x\,) \,.
\end{align*}
The computation of the reference solution $u^{\, \mathrm{ref}}\, (\,x \,, t \,)$ is detailed in further Sections.


\section{Linear transfer in porous material}
\label{sec:AN1}

The first case of linear moisture transfer is considered from \cite{Rode2006, Bednar2005} to analyze the moisture buffer effects in  a $500$-$\mathsf{mm}$ aerated concrete under isothermal condition, at a temperature of $23\, \unit{^{\circ}C}$. The vapor permeability is $\dm \,=\, 3 \cdot 10^{\,-11}\, \unit{s}$ and its moisture storage is $\cm \,=\, 1.85 \cdot 10^{\,-4}\, \unit{kg/m^3/Pa}$ \cite{Bednar2005}. The uniform initial vapor pressure in the material is $\Pvi \,=\, 842\, \unit{Pa \,}$, corresponding to a relative humidity of $30\,\%\,$. The total time of simulation corresponds to $120\, \unit{h}$. The left boundary is set to a constant vapor pressure, identical to the initial condition. At the right boundary, the relative humidity varies sinusoidally between $33\,\%$ and $75\,\%\,$, with a period of $24\, \unit{h}$. The convective vapor transfer coefficient is set to $2 \cdot 10^{\,-8}\, \unit{s/m}$.

\begin{figure}
  \centering
  \subfigure[a][\label{fig:precase_mbv}]{\includegraphics[width=0.48\textwidth]{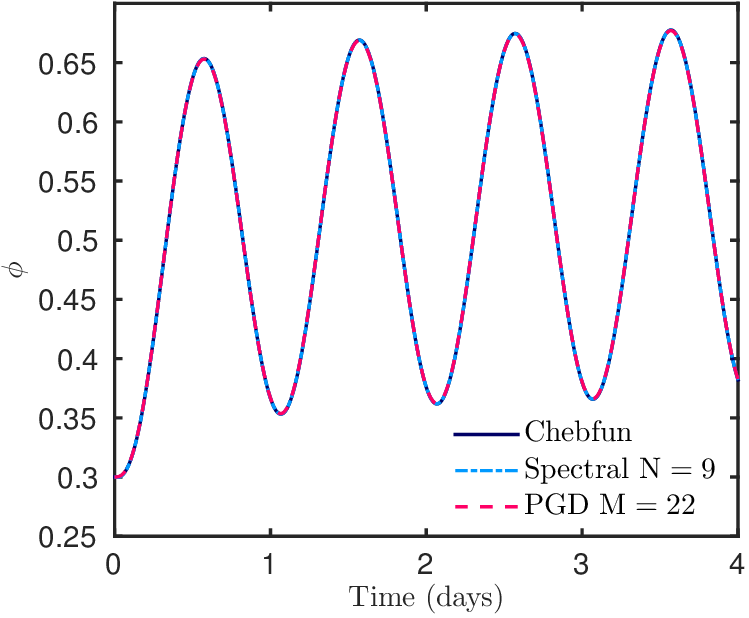}}
  \subfigure[b][\label{fig:precase_profil}]{\includegraphics[width=0.49\textwidth]{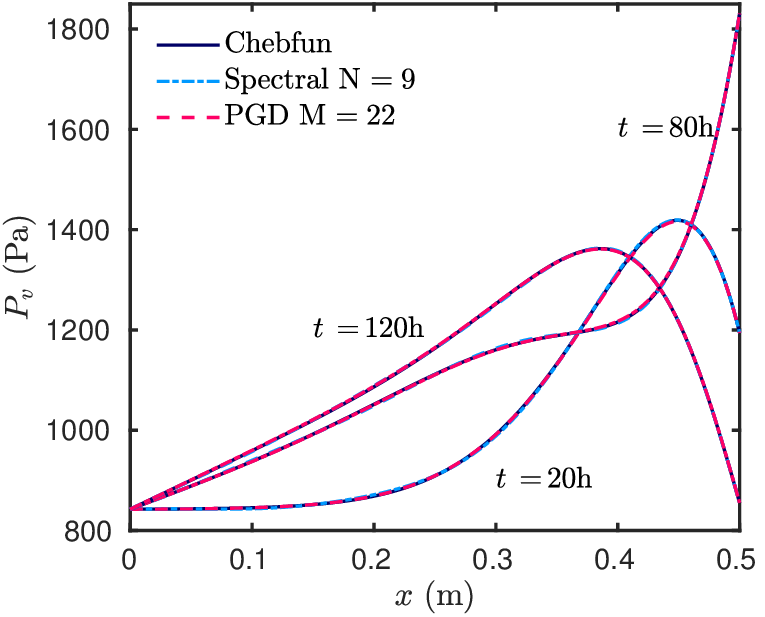}}
  \caption{\small\em (a) Relative humidity at the right boundary as function of time, and (b) pressure vapor profiles for $t = \{20\,,\,80\,,\,120\}\,\mathsf{h}\,$.}
\end{figure}

The solution of the problem has been computed for a discretization of $\Delta \xs \egal 5 \cdot 10^{\,-3}$ and $\Delta \ts \egal 10^{\,-1}\,$. The physical phenomena are well represented, as illustrated in Figure~\ref{fig:precase_mbv} with the time evolution of the relative humidity at $x\,=\,0.47\, \mathsf{m}$. The variations follow the ones of the right boundary conditions and, the diffusion process goes towards the periodic regime. It can be noted a good agreement between the two reduced-order models. Furthermore, the vapor pressure profile is shown in Figure~\ref{fig:precase_profil} for $t \,=\, \{ 20\,,\,80\,,\,120\}\, \mathsf{h}$, enhancing the good accuracy of the solution to represent the physical phenomenon.

\begin{figure}
  \centering
  \subfigure[a][\label{fig:precase_error}]{\includegraphics[width=0.48\textwidth]{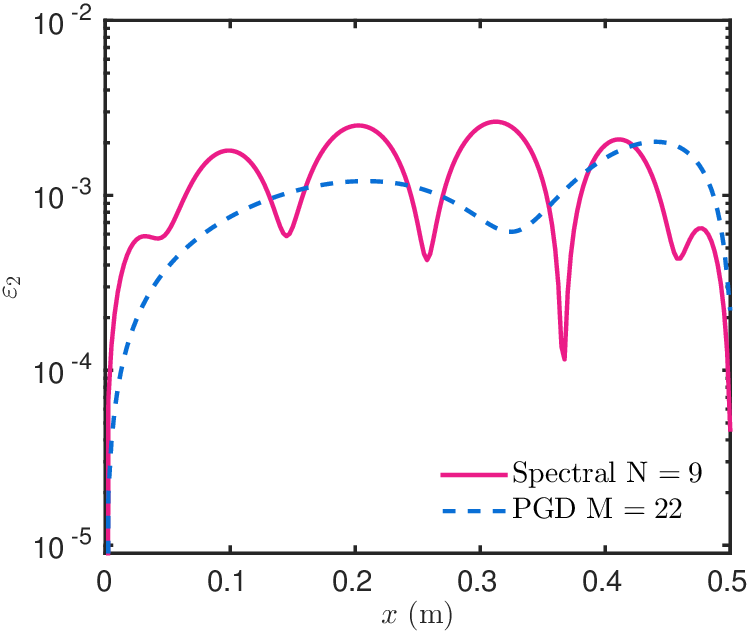}}
  \subfigure[b][\label{fig:precase_error_modes}]{\includegraphics[width=0.48\textwidth]{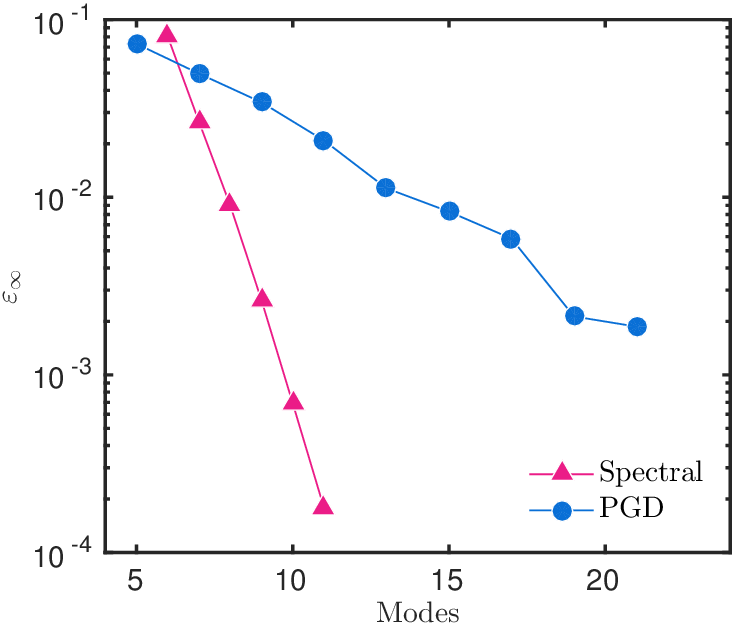}}
  \caption{\small\em (a) Error as a function of $x\,\mathsf{(m)}$ and (b) error as a function of the number of modes.}
\end{figure}

\begin{figure}
  \centering
  \subfigure[a][\label{fig:precase_profil_PGD}]{\includegraphics[width=0.48\textwidth]{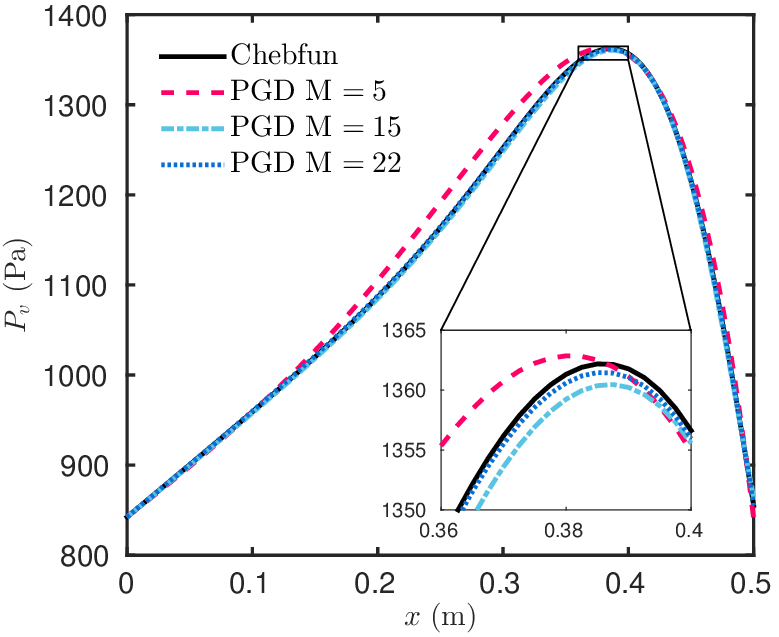}}
  \subfigure[b][\label{fig:precase_profil_SPEC}]{\includegraphics[width=0.48\textwidth]{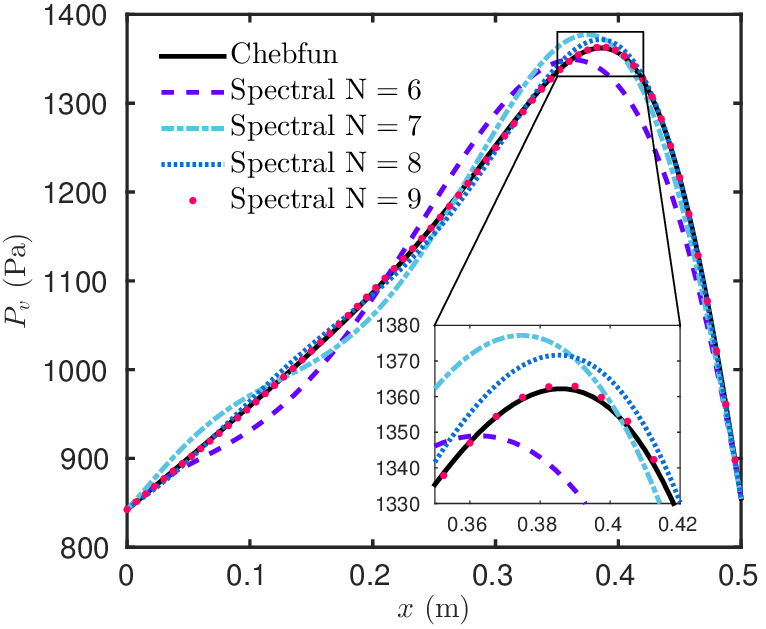}}
  \caption{\small\em Vapor pressure profiles for different number of modes (a) for the PGD and (b) for the Spectral--ROM.}
\end{figure}

The absolute error of the reduced-order model is of the order of $\O \simeq (\,10^{\,-4}\,)$, as illustrated in Figure~\ref{fig:precase_error}. The methods are compared with a reference solution, which has been computed using the \texttt{Matlab} open source toolbox \texttt{Chebfun} \cite{Driscoll2014}. To give a solution with this order of accuracy, the PGD needed $22$ modes, while the Spectral only required $9$ modes. It corresponds to $21$ and $7$ degrees of freedom, respectively. Figure~\ref{fig:precase_error_modes} presents the error $\varepsilon_{\,\infty}$ as a function of the number of modes. As we increase the number of modes, the solution of the Spectral--ROM converges faster than the PGD solution because the convergence of the Spectral method is exponential. To illustrate the convergence of the solution, the profile of the vapor pressure for the last time instant of simulation is represented with different numbers of modes in Figures~\ref{fig:precase_profil_PGD} and \ref{fig:precase_profil_SPEC} for the PGD and the Spectral--ROM, respectively.

\begin{figure}
  \centering
  \subfigure[a][\label{fig:precase_spc_A}]{\includegraphics[width=0.53\textwidth]{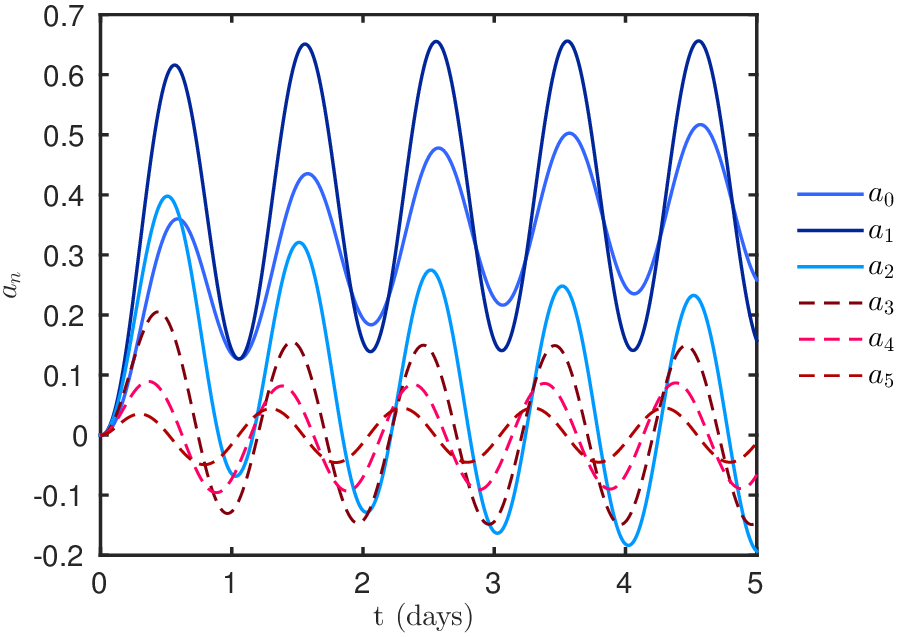}}
  \subfigure[b][\label{fig:precase_spec_eigen}]{\includegraphics[width=0.44\textwidth]{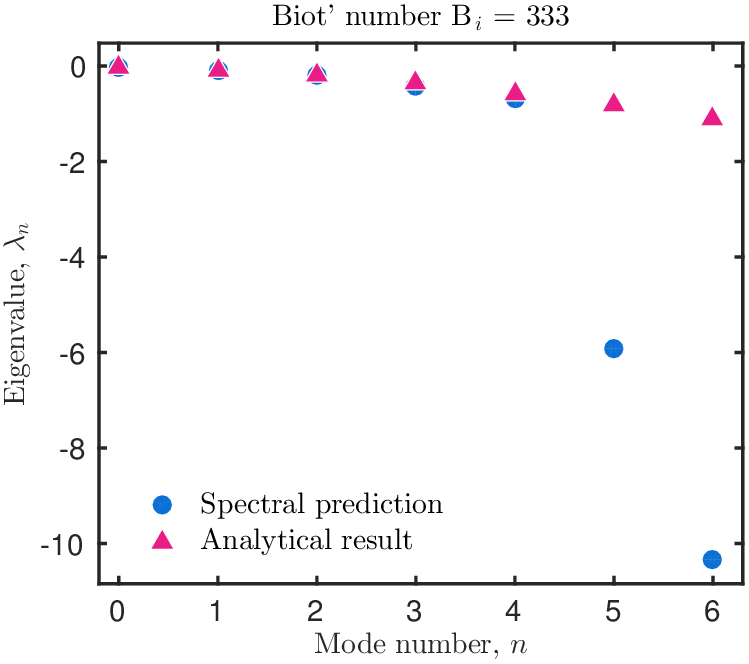}}
  \caption{\small\em (a) Spectral coefficients as sets of time and, (b) Eigenvalues of the Analytical and of the Spectral solution corresponding to the first modes.}
\end{figure}

\begin{figure}
  \centering
  \includegraphics[width=0.48\textwidth]{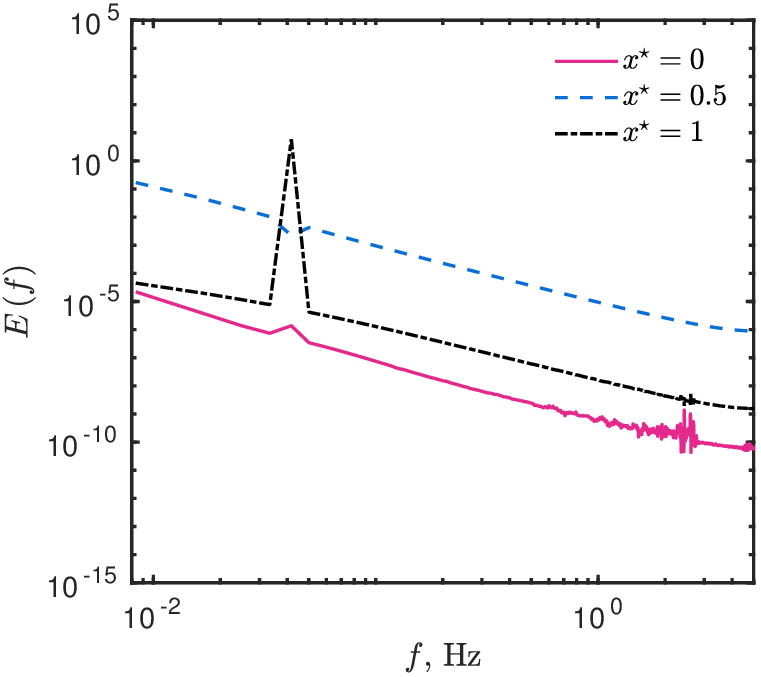}
  \caption{\small\em \textsc{Fourier} power spectrum of the Spectral solution computed in the boundaries and in the middle of the material $\xs \in \left\lbrace 0\,,\, 0.5\,,\, 1 \right\rbrace\,$.}
  \label{fig_AN1:FPS}
\end{figure}

The reduced system of ODEs, of size $7$ is implemented in \texttt{Matlab}, and the Spectral coefficients $\{a_{\, n}\,(\,t\,)\}$ are calculated for any intermediate time instants by the solver \texttt{ODE45} \cite{Shampine1997}. These Spectral coefficients are shown in Figure~\ref{fig:precase_spc_A}. It can be seen that the first coefficients have the highest magnitude making the solution to converge with a few modes (an order of $10$), which happens thanks to the fact that the \textsc{Chebyshev} polynomials have excellent approximation properties for smooth function. In addition, the last coefficient determines the magnitude of the residual, implying that the error will not be lower than the magnitude of the last coefficient $a_{\,n}$, as explained in \cite{Gasparin2018}. A brief comparison with an analytical solution, built on \textsc{Fourier} decomposition \cite{Ozisik1993}, reveals that the eigenvalues of the Spectral method decrease faster, as shown in Figure~\ref{fig:precase_spec_eigen}. Note that eigenvalues of the analytical solution do not have to coincide with the ones of the Spectral method since the basis functions are not the same for the \textsc{Chebyshev} polynomials and the trigonometric ones. Figure~\ref{fig_AN1:FPS} shows the \textsc{Fourier} power spectrum function of the signal frequency per unit of time, generated by the fast \textsc{Fourier} transform. In this Figure, one peak is observed in the signal frequency which is consistent with the step of the relative humidity on the right boundary.

\begin{figure}
\centering
\subfigure[a][\label{fig:precase_pgd_F}]{\includegraphics[width=0.48\textwidth]{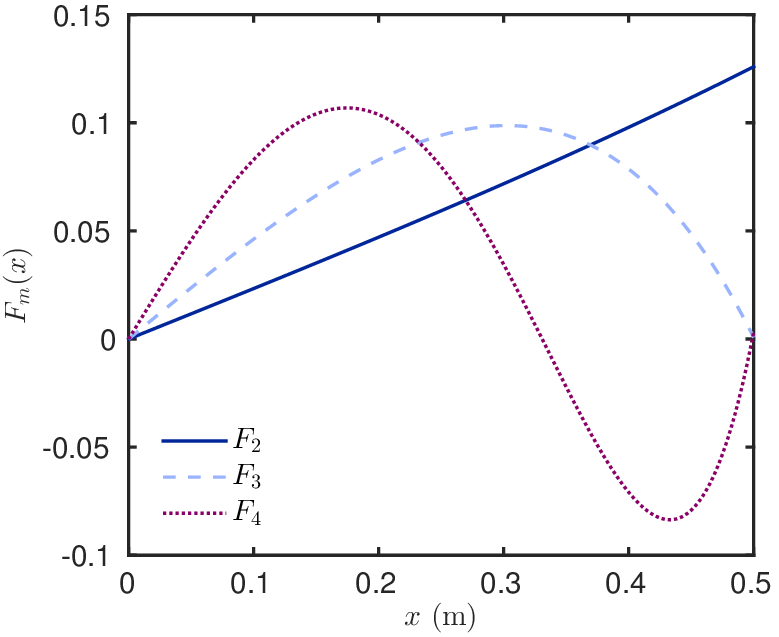}}
\subfigure[b][\label{fig:precase_pgd_G}]{\includegraphics[width=0.48\textwidth]{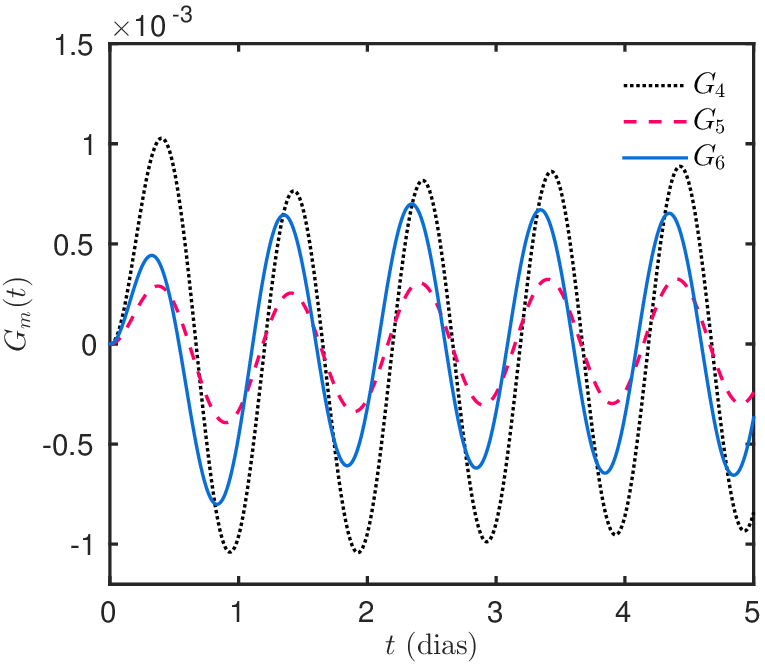}}
\caption{\small\em (a) $F_m$ tensor function; (b) $G_m$ tensor function.}
\end{figure}

Regarding the PGD, Figures~\ref{fig:precase_pgd_F} and \ref{fig:precase_pgd_G} present the first modes, depending on time and space, respectively. They do not have a physical meaning and constitute a separated representation of the solution. Their tensorial product enables to compute the solution of the problem. It is a similar approach for the Spectral--ROM, where the coefficients $a_{\,n}\,(\,t\,)$ are multiplied by the \textsc{Chebyshev} polynomials.

\paragraph{Remarks on \emph{a posteriori} POD method}~\\

\begin{figure}
  \centering
  \subfigure[a][\label{fig:precase_pod}]{\includegraphics[width=0.48\textwidth]{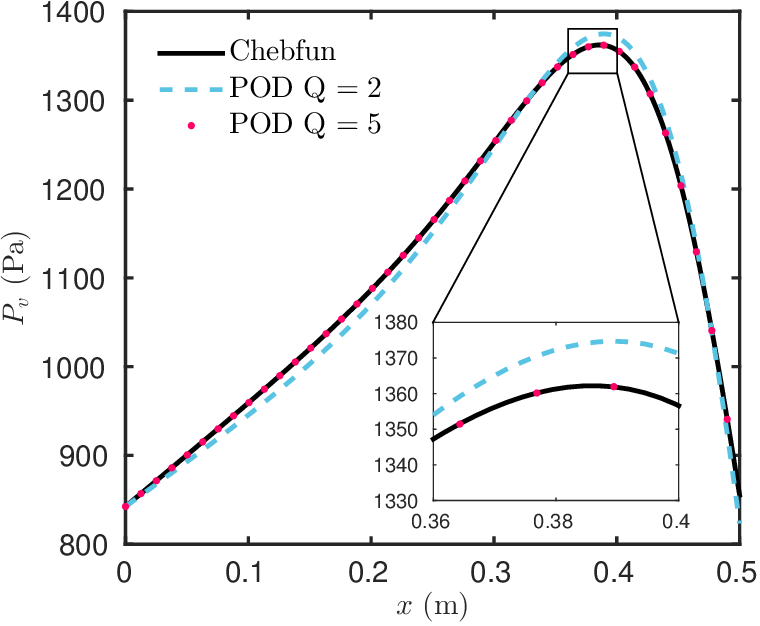}} \hspace{0.1cm}
  \subfigure[b][\label{fig:precase_pod2}]{\includegraphics[width=0.48\textwidth]{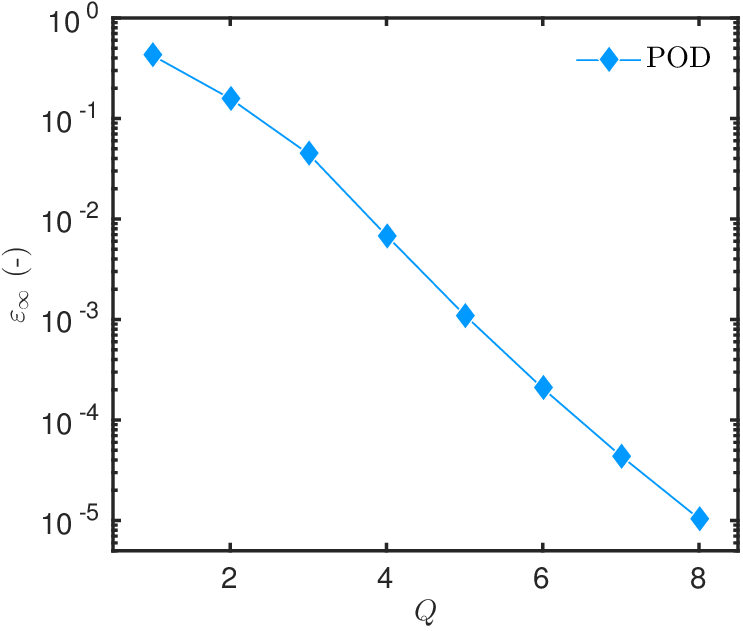}}
  \caption{\small\em Vapor pressure profiles for different number of modes computed with the POD (a) and the error $\varepsilon_{\,\infty}$ as a function of the number of modes $Q$ (b).}
\end{figure}

Here, the purpose is to compare the PGD and Spectral model order reduction to the well-established POD approach. For this, the reference solution $u$ obtained with the \texttt{Matlab} open source toolbox \texttt{Chebfun} was used to compute the correlation matrix $\textsc{R}\,$, with elements $\{r_{\,ik}\}$ given by:
\begin{align*}
  r_{\, ik} \egal \sum_{j\,=\,1}^{N_{\,t}} u \, (\, \,x_{\,i} \,,\, t_{\,j} \,) \; u \, (\, \,x_{\,k} \,,\, t_{\,j} \,) \,.
\end{align*}
Then, the singular values $\lambda$ of the correlation matrix are computed. A truncation is operated in the eigenvectors basis to define $\phi_{\,i}$ the spatial basis function composed of the $Q$ eigenvectors of the correlation matrix $r\,$. Thus, the solution of the POD reduced-order model (POD--ROM) approximates the solution of the problem by:
\begin{align*}
  u\, \, (\,x\,,t\,) \ \approx\ u_{\, Q}\, (\,x\,,t\,) \eqdef \sum_{i\, =\, 0}^Q \, a_{\,i}\, (\,t\,)\, \phi_{\,i}\, (\,x\,)\,,
\end{align*} 
where $Q$ is the number of modes corresponding to the model order reduction. For this case study, Figure~\ref{fig:precase_pod} shows the convergence of the solution obtained with a POD reduced-order model. Moreover, Figure~\ref{fig:precase_pod2} presents the error with the reference solution as a function of the modes $Q\,$. The solution of the POD--ROM converges faster than the PGD and Spectral approaches. Only $Q \egal 5$ modes are sufficient to compute a solution accurate to the order $\O\,(10^{\,-3})\,$, which is lower than the one for the PGD or Spectral--ROM for the same accuracy. However, as underlined in the Introduction section, the POD approach is \emph{a posteriori}. To build the POD reduced-order model, a preliminary computation of the solution was required, which is a non-negligible restriction.

\section{Computing parametric solution using reduced-order models}
\label{sec:AN2}

ding performance assessment. Consider, for instance, the analysis of the wall behavior in terms of heat and mass transfer, as a function of different parameters such as thermal inertia, vapor permeability, insulation thickness, among others. In the context of environmental issues and thermal regulations, the wall behavior may be optimized as a function of those parameters. Several studies of parametric simulations are presented in the literature. In \cite{Axaopoulos2014, Ozel2011, Al-Sanea2005, Ibrahim2015, Bond2013, Al-Sanea2011, Yuan2016}, numerical methods are used to determine the optimum insulation thickness of different wall configurations. In \cite{Aste2009}, the influence of wall thermal inertia on the energy consumption was investigated by using the \texttt{EnergyPlus} program for $24$ construction types. In \cite{Labat2016}, the MBV of five hemp concrete materials were assessed using a numerical method. Those parametric simulations are based on models using numerical methods due to almost no restriction in terms of boundary conditions, geometry, material properties, among other considerations. Nevertheless, for parametric studies, they require large numbers of simulations. Indeed, the numerical model is not dependent on the parameters of interest. Thus, a computation of the numerical model is required for each value of the parameters within their domain of variation, demanding a high calculation cost, even after the dramatic evolution of computer hardware since the $1970$'s. Therefore, reduced-order model can be used to perform efficient parametric studies with a limited computational costs.


\subsection{Extension of model reduction techniques to parametric problems}

The issue of solving a parametric problem is to compute the solution $u$ of Eq.~\eqref{eq:heat1d} depending on the usual time and space coordinates $x$ and $t$ as well as the parameter $\nu$ of the problem. Thus, the solution is seek as $u\, (\, x\,, t\,; \nu\,)\,$, where $\nu$ is defined as a coordinate of the problem within a given interval $\nu \ \in \ \bigl[\, \nu_{\,\mathrm{min}} \,,\, \nu_{\,\mathrm{max}} \, \bigr] $. Here we note $N_{\nu}$ the number of elements (cardinal) of the domain $\bigl[\, \nu_{\,\mathrm{min}} \,,\, \nu_{\,\mathrm{max}} \, \bigr]\,$.


\subsubsection{Spectral reduced-order model}

For the parametric study, we want to compute the solution as a function of $(\, x \,, t \,; \nu\, )\,$:
\begin{align*}
  u\, (\, x\,, t\,; \nu\,)\ \approx \ u_{\,n}\, (\, x\,, t\,; \nu\,) \egal \sum_{n\,=\,0}^{N} \, a_{\,n}\, (\, t\,; \nu\,)\ \phi_{\,n}\, (\,x\,)\,.
\end{align*}
The basis function will always depend only on $x$, that is why the parameter $\nu$ is calculated with Spectral coefficients $\{a_{\,n}\,(\, t\,; \nu\,) \}\,$. As it is not straightforward to compute the coefficients $a_{\,n}\, (\, t \,;\, \nu\, )$ depending on both parameters, we compute the solution for each value of the parameter $\nu\,$, using a loop. The latter can easily be parallelised on high-performance computer systems. It would be possible to vectorize the computation of the parametric Spectral--ROM solution although the method would significantly loose its speed calculation.


\subsubsection{PGD reduced-order model}

To compute a parametric solution of Eq.~\eqref{eq:heat1d} $u\, (\, x \,, t \,; \nu \,)\,$, the PGD approach assumes a separated tensorial representation of the solution: 
\begin{align*}
  u \, (\, x \,, t \,; \nu \,) \ \approx \ u_{\,m}\, (\, x \,, t \,; \nu \,) \egal \sum_{m\;=\;1}^{M} F^{\,m} \, (\, x \,) \ G^{\,m} \, (\, t \,) \ H^{\,m} \, (\, \nu \,) \,.
\end{align*}
Functions $\bigl(\, F \,,\, G \,,\, H \,\bigr)$ are computed following the methodology described in Section~\ref{sec:PGD}. Interested readers may refer to \cite{Chinesta2011, Chinesta2011a, Berger2017d} for complementary details on the methodology. The important point, is that the solution $u$ is computed \emph{at once} as a function of the coordinates $x\,$, $t$ and $\nu\,$.


\subsection{Case study}

For this case, we seek for a parametric solution of problem Eq.~\eqref{eq:moisture_dimensionlesspb_1D}. The vapor pressure is computed as a function of time $t$, space $x$ and moisture storage capacity $\cm$. As in the previous case, simulations are preformed in order to reproduce experiments that estimate the moisture buffer value of the materials. Thus, the right boundary condition is exposed to cyclic changes of relative humidity between $33\, \unit{ \% }$ and $75\, \unit{ \% }$, with a $24\; \unit{h}$ period. The total time of simulation is still $120\;\unit{h}$. The convective vapor coefficient is $\hv \egal 2 \cdot 10^{\,-8}\;  \unit{s/m}$. The left boundary is set to a constant vapor pressure, identical to the initial condition $\Pvi \egal 842\; \unit{Pa \,}$. Simulations undergo at a constant $23 \unit{^{\circ} C}$ temperature. All $500\text{-}\unit{mm}$ materials have the same vapor permeability, $\dm \egal 2.4 \cdot 10^{\,-11}\; \unit{s}$, while the moisture storage capacity varies in the interval $\Omega_{\,c} \egal 1.2 \cdot 10^{\,-3}$ and $6 \cdot 10^{\,-3}\; \unit{kg/m^3/Pa}$ \cite{Rode2006}.

First, we perform a simulation for $10$ different values of moisture storage capacity in the interval $\Omega_{\,c}$, representing different kinds of materials. Two different techniques of reduction order models were employed, the PGD and the Spectral--ROM. To validate these methods, results were compared to the reference solution, constructed with the \texttt{Matlab} open source toolbox \texttt{Chebfun}.

\begin{figure}
  \centering
  \subfigure[a][\label{fig:case1_massall}]{\includegraphics[width=0.48\textwidth]{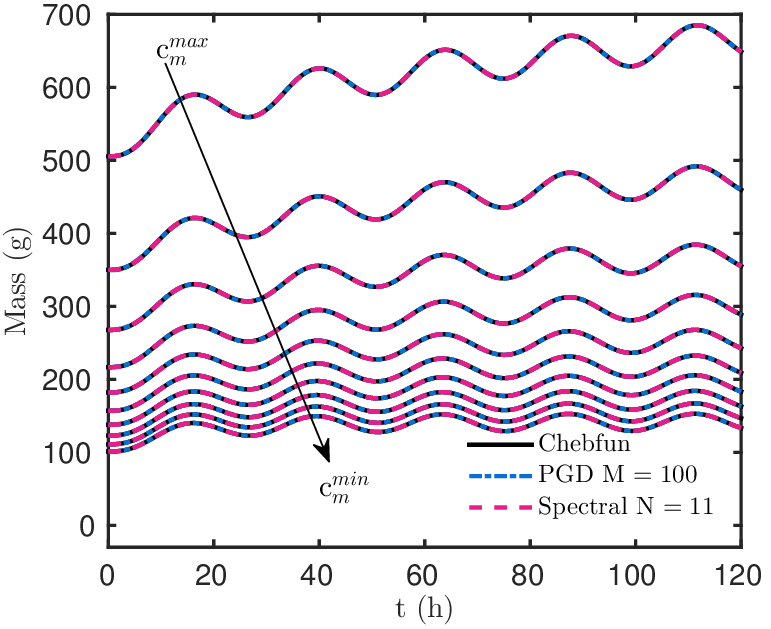}}
  \subfigure[b][\label{fig:case1_massa2first}]{\includegraphics[width=0.48\textwidth]{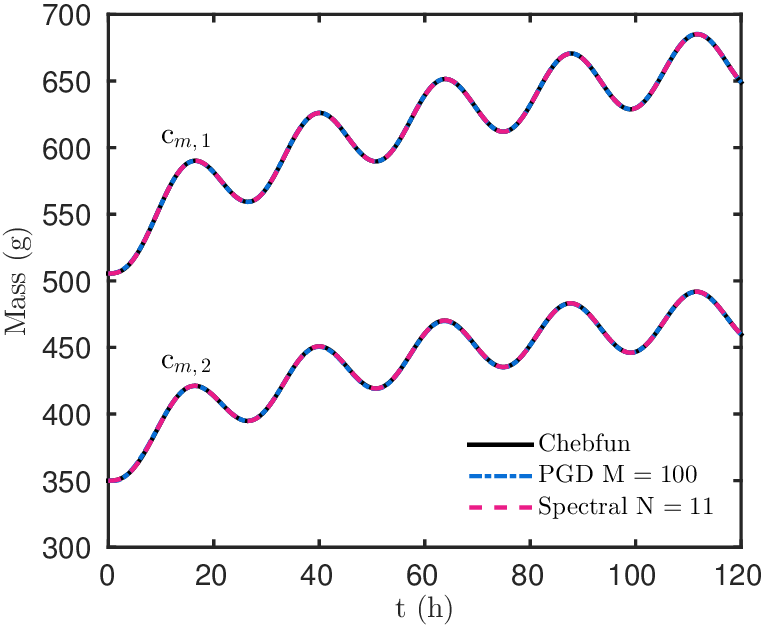}}
  \caption{\small\em (a) Water content increase for all values of parameter $\cm$, and (b) Water content increase for the two highest values of moisture storage capacity.}
\end{figure}

\begin{figure}
  \centering
  \subfigure[a][\label{fig:case1_last_frofile}]{\includegraphics[width=0.49\textwidth]{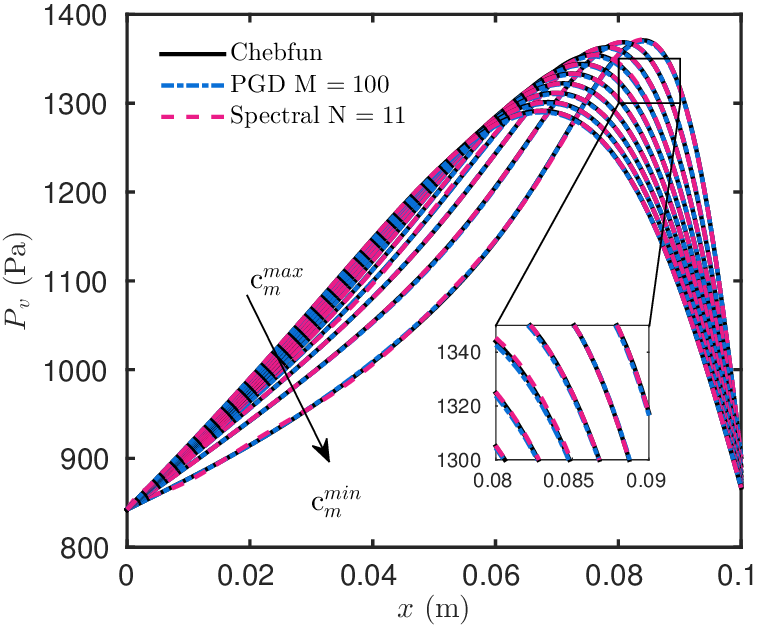}} \hspace{0.1cm}
  \subfigure[b][\label{fig:case1_error}]{\includegraphics[width=0.48\textwidth]{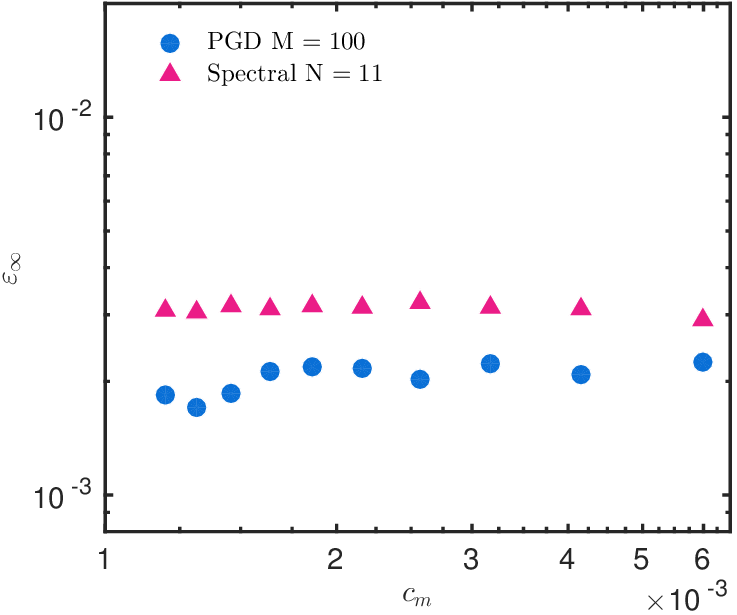}}
  \caption{\small\em (a) vapor pressure profiles at $t\ =\ 120\,\mathsf{h}$, for $\cm \in [1.2 \cdot 10^{\,-3};\, 6 \cdot 10^{-3}]$ $\mathsf{kg/m^3/Pa}$, and (b) the error in function of all values of moisture storage capacity between this interval.}
\end{figure}

Figure~\ref{fig:case1_massall} shows the mass content of each material among the simulation time. Even with a low difference between the highest and the lowest values of storage capacity, it is possible to observe significant variations of the weight as the the storage capacity increases, retaining more moisture. Furthermore, Figure~\ref{fig:case1_massa2first} presents the weight for the two highest values of moisture storage capacity, corresponding to $6\cdot 10^{\,-3}$ and $4.15\cdot10^{\,-3}$ $\mathsf{kg/m^3/Pa}$. The last profile of the pressure vapor for all values of moisture storage capacity is represented in Figure~\ref{fig:case1_last_frofile}. In these figures, the PGD and the Spectral--ROM are in a good agreement with the reference solution.

\begin{figure}
  \centering
  \includegraphics[width=0.69\textwidth]{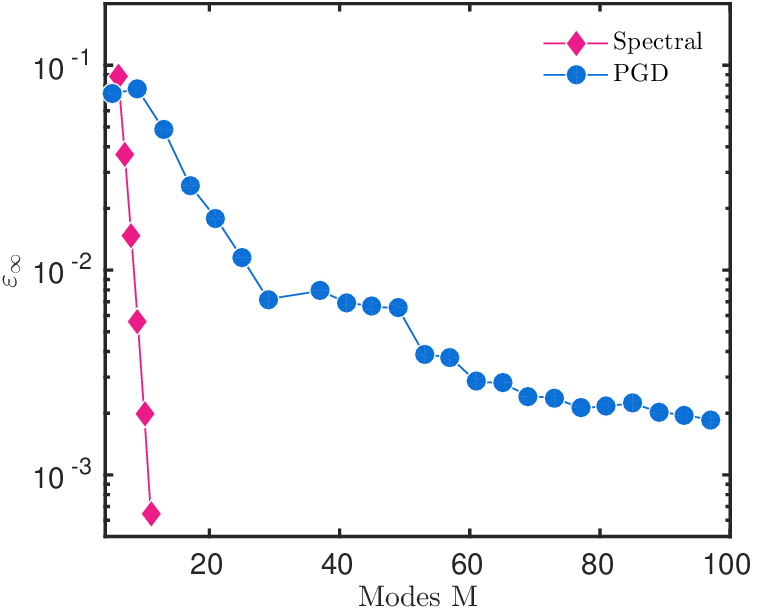}
  \caption{\small\em Error calculated for $\cm\,=\, 1.46 \cdot 10^{-3}\,\mathsf{kg/m^3/Pa}$ in function of the number of modes M.}
  \label{fig:case1_error_modes_PGD}
\end{figure}

To compute the parametric study, with the same order of accuracy, the PGD needed around $M\ =\ 100$ modes, while the Spectral--ROM used only $N\ =\ 11$ modes. This difference comes from the nature of the methods, that are constructed by different ways. The error $\varepsilon_{\,\infty}$ is shown as a function of the storage capacity values in Figure~\ref{fig:case1_error}. The methods were constructed in order to give the same order of accuracy, around $\O\,(\,10^{\,-3}\,)\,$. It should be noted that the Spectral--ROM can give more accurate results, with the same degrees of freedom, by increasing the tolerance in the \texttt{ODE45} \texttt{Matlab} function, when the reduced system is being computed. The degrees of freedom of the Spectral--ROM were predetermined based on the previous case and by the order of the parameters values. Meanwhile, the PDG computes its solution if the residual is lower than a given tolerance. The rate of convergence of the PGD approach is illustrated in Figure~\ref{fig:case1_error_modes_PGD}, presenting the error as a function of the number of modes.

The numbers of operation for each approach can be estimated, remembering $N_{\,x}$ and $N_{\,t}$ stand for the number of elements according to the discretization of the space and time domains, respectively. The quantity $N_{\,\nu}$ represents the number of elements of parameter $\cm$ considered for the parametric study. A standard approach based on implicit \textsc{Euler} schemes requires $N_{\,x} \cdot N_{\,t} \cdot N_{\, \nu}$ operations. Considering the discretization parameters to reach the given accuracy $N_{\,t} \egal 1.2 \cdot 10^{\,5}$ and $N_{\,x} \,=\, 4 \cdot 10^{\,2}\,$, the number of operations scales with \cite{Gasparin2017}: 
\begin{align*}
  & \text{\textsc{Euler} implicit:} && \O\simeq\ \Bigl(\, N_{\,x} \cdot N_{\,t} \cdot N_{\, \nu} \, \Bigr)\ \simeq \ \Bigl(\, 4.8 \cdot 10^{\, 7} \cdot N_{\, \nu} \, \Bigr) \,.
\end{align*}

For the Spectral--ROM, the number is related to the solution of the system of ODEs~\eqref{eq:system_ODE}, computed in this case with the \texttt{Matlab} \texttt{ODE45} solver. It is based on the iterative \textsc{Runge}--\textsc{Kutta} method to approximate the solution. The number of operation depends on the tolerance of the solver, which has a maximum tolerance of $ \sim 10^{\,-5}$ for \texttt{ODE45}. Thus, we have: 
\begin{align*}
  N_{\,t} \ \simeq \ \dfrac{\tau}{\Delta t} \ \simeq \ \dfrac{\tau}{(\mathsf{tol})^{\,1/5}} \,,
\end{align*}
where $T$ is the total time of simulation. At each time step, the \textsc{Runge}--\textsc{Kutta} needs to compute six times the vector product $\A_{\, n \times n}$, where $n$ depends on the degree of freedom $N$ of the solution ($n \egal N \moins 2$). Thus, it leads to $6\cdot n^{\, 2}$ operations to perform, knowing that $n$ scales with $10\,$. Consequently, the total number of operations for the Spectral--ROM scales with:
\begin{align*}
  \O\,\biggl(\, \dfrac{6 \cdot (N \moins 2)^{\,2} \cdot \tau}{(\mathsf{tol})^{\,1/5}}\, \biggr) \,.
\end{align*}
For this parametric case, knowing that the tolerance was set to $10^{\, -3}\,$, with $N\ \simeq\ 11$ modes, the number of operations performed by the Spectral--ROM is expressed as:
\begin{align*}
  \text{Spectral--ROM:}\quad  \O\,\biggl(\, \dfrac{6 \cdot (11 \moins 2)^{\,2} \cdot \tau}{(\mathsf{10^{\,-3}})^{\, 1/5}} \cdot N_{\, \nu}  \, \biggr)\ &\simeq\ \Bigl(\, 2.3 \cdot 10^{\, 5}\cdot  N_{\, \nu} \, \Bigr) \,.
\end{align*}

For the PGD, the number of operations depends on the number of modes $M\,$, the number of iterations $S$ required for the fixed point algorithm to converge and the sum of the number of spatial, temporal and parameters elements $N_{\,x}\,$, $N_{\,t}$ and $N_{\, \nu}\,$, respectively \cite{Berger2016c}. For this application, the PGD fixed point algorithm requires around $S \egal 20$ iterations at each mode and $M \egal 100$ modes. The discretization parameter used for the PGD are $N_{\,t} \egal 1200$ and $N_{\,x} \egal 100\,$. Therefore, the number of operations for the PGD approaches scales with:
\begin{align*}
  \text{PGD:}\quad \O\,\Bigl(\, S \cdot M \cdot (N_{\,t} \plus N_{\,x} \plus N_{\, \nu} ) \, \Bigr)\ &\simeq\ \Bigl(\, 2.6 \cdot 10^{6} \plus 2000 \cdot N_{\, \nu} \, \Bigr) \,.
\end{align*}

\begin{figure}
  \centering
  \includegraphics[width=0.68\textwidth]{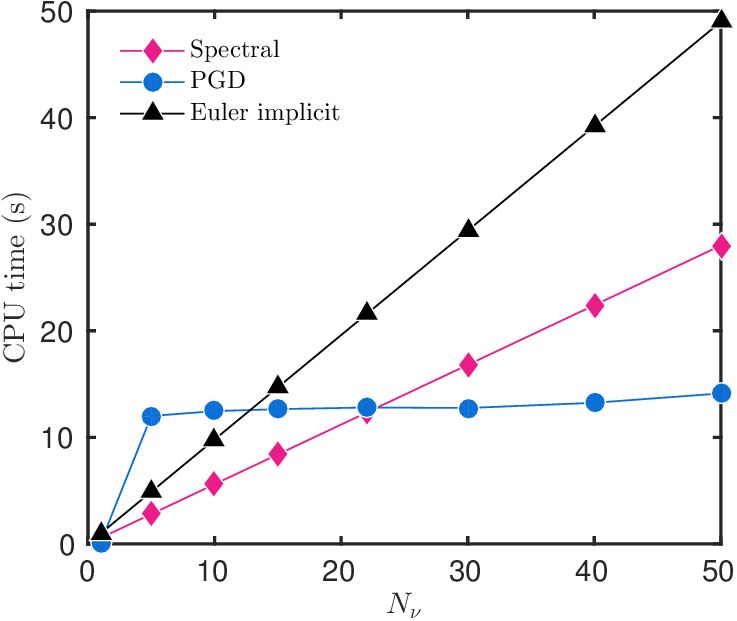}
  \caption{\small\em CPU time $\unit{(s)}$ as a function of the number of elements $N_{\,\nu}$ of parameter $\cm$.}
  \label{fig:case1_cpu_time}
\end{figure}

Therefore, the number of operations increases faster for the Spectral--ROM and for the standard \textsc{Euler} methods than for the PGD approach. After some kind of initial investment, the increase in the number of operations for the PGD is much slower than for other approaches. The advantage of the PGD, in this case, it is related to its ability to compute \emph{at once} the solution depending on the three coordinates, whereas the Spectral--ROM computes the solution for each value of moisture storage capacity independently, by a loop. It should be noted that the \textsc{Euler} approach, based for instance on backward time centered space, is a low order approximation of the solution, providing a less accurate solution than the Spectral--ROM. An interesting advantage of the PGD is the low storage cost of the solution thanks to the tensorial representation of the solution. This feature may be particularly interesting for real time applications. These features also impact the CPU time of each algorithm, which has been evaluated using \texttt{Matlab} platform on a computer with Intel i$7$ CPU and $8$GB of RAM. Figure~\ref{fig:case1_cpu_time} shows the CPU time as a function of the number of elements of the parameter $\cm$. For comparison, the CPU time required using the \textsc{Euler} implicit scheme is also reported. Since the Spectral--ROM has a reduced system to solve, its computational time drops significantly when compared to traditional methods. However, the loop to simulate the parametric study increases the CPU time linearly with the number of elements. For few numbers of elements of parameter $\cm$, around $20\,$, the Spectral--ROM is faster than the PGD. Yet, if the number of elements increases, the PGD is a more attractive method. In addition, the large original model, based on implicit \textsc{Euler} scheme, requires an important extra CPU time to compute the parametric solution.


\section{Nonlinear transfer in porous material}
\label{sec:AN3}

The last case considers nonlinear transfer with moisture-dependent material properties. Therefore, the diffusion coefficient $\nu$ depends on the field $u$ and Eq.~\eqref{eq:heat1d} becomes: 
\begin{align}\label{eq:heat1d_NL}
   \pd{u}{t} \egal \div \Bigl(\, \, \nu \, (\,u\,) \grad u \, \Bigr) \,.
\end{align}


\subsection{Extension of model reduction techniques to nonlinear problems}

Some adaptations of the methods must be carried out in order to consider nonlinear diffusion transfer. For this, in what follows the reduced-order models are described.


\subsubsection{Spectral reduced-order model}

In order to apply better the Spectral method, Eq.~\eqref{eq:heat1d_NL} is written in the non-conservative form:
\begin{align}\label{eq:rearang_diff_nonlin}
  \pd{u}{t} &\egal \nu \, (\, u \, ) \, \pd{^{\,2} u}{x^{\,2}} \plus \lambda \, (\, u \, )\, \pd{u}{x} \,,
\end{align}
where,
\begin{align*}
  \lambda \, (\, u \,) &\eqdef  \dfrac{\mathrm{d} \Bigl(\nu \, (\, u \,)\Bigr)}{\mathrm{d} x}\,.
\end{align*}
By using Spectral methods the unknown $u\,(\,x\,,\,t\,)$ is approximated by the finite sum~\eqref{eq:series_ap} and, the derivatives can be written so that the \textsc{Chebyshev} polynomials remain the same, as in the linear case of Eq.~\eqref{eq:derivatives}. Thus, Eq.~\eqref{eq:rearang_diff_nonlin} becomes:
\begin{align}\label{eq:rearang_diff_nonlin2}
  \sum^n_{i\, = \, 0}\, \dot{a}_{\,i}\, (\, t \,)\, T_{\,i}\, (\, x \,) \egal \nu \, \Biggl(\, \sum_{i\, = \, 0}^n\, a_{\,i}\, (\, t \,)\, T_i\, (\, x \,) \, \Biggr) \, \sum_{i\, =\, 0}^n \Tilde{\Tilde{a}}_{\,i}\, (\, t \,)\, T_{\,i}\, (\, x \,) \plus \nonumber \\ 
  \lambda \, \Biggl( \, \sum_{i\, = \, 0}^n a_{\,i}\, (\, t \,)\, T_{\,i}\, (\, x \,)\,  \Biggr) \sum_{i\, =\, 0}^n \tilde{a}_{\,i}\, (\, t \,)\, T_{\,i}\, (\, x \,) \,.
\end{align}
The nonlinear terms $\nu\, \Bigl(\, \sum_{i\, = \, 0}^n\, a_i\, (\, t \,)\, T_i\, (\, x \,) \, \Bigr)$ and $\lambda\, \Bigl( \, \sum_{i\, = \, 0}^n\, a_{\,i}\, (\, t \,) \, T_{\,i}\, (\, x \,)\,  \Bigr)$ are treated by applying the \textsc{Tau}--\textsc{Galerkin} method and the \textsc{Chebyshev--Gau}\ss{} quadrature \cite{Gasparin2018}.
Contrary to the linear case, the boundary conditions cannot provide an explicit expression for the two last coefficients $a_{\,n}\, (\, t \,)$ and $a_{\,n-1}\, (\, t \,)\,$. Thus, it is not possible to compute the solution in the same way. Although, with all elements listed before, it is possible to set the system to be solved by adding two algebraic expressions for the boundary conditions. It results in a system of Differential-Algebraic Equations (DAEs) with the following form:
\begin{align*}
  \M \, \dot{a}_{\,n}\, (\, t \,) \egal \A \, a_{\,n}\, (\, t \,) \plus \b\, (\, t \,) \,,
\end{align*}
where, $\M$ is a diagonal and singular matrix ($\mathrm{rank}\, (\,\M\,)\,=\,n\ -\ 2$) containing the coefficients of the \textsc{Chebyshev} weighted orthogonal system, $\b\, (\, t \,)$ is a vector containing the boundary conditions and, $\A \cdot a_{\,n}\, (\,t\,)$ is composed by the right member of Eq.~\eqref{eq:rearang_diff_nonlin2}.
The initial condition is given by Eq.~\eqref{eq:system_ODE_int} and the DAE system is solved by \texttt{ODE15s} or \texttt{ODE23t} from \texttt{Matlab}. For further details, interested readers may consult \cite{Gasparin2018, Shampine1997}.


\subsubsection{PGD reduced-order model}

To treat the nonlinearity of the problem, at the enrichment step $m \ < \ M\,$, the nonlinear term $\nu\,(\,u\,)$ is approximated using the solution from previous steps:
\begin{align*}
  \nu \, (\,u\,) \egal \nu \, \Biggl(\, \sum_{i\,=\,1}^{M}\, F^{\,i} \, (\, x \,) \ G^{\,i} \, (\, t \,)  \plus R\, (\, x \,) \ S\, (\, t \,) \,\Biggr)\ \simeq \ \nu \, \Biggl(\, \sum_{i\,=\,1}^{M}\, F^{\,i} \, (\, x \,) \ G^{\,i} \, (\, t \,)  \,\Biggr) \,.
\end{align*}
Then, the matrix of the coefficient $\nu$ is separated into a tensorial product in the space and time directions, using a \emph{Singular Value Decomposition} (\svd) \cite{Golub1996} or a \emph{Discrete Empirical Interpolation Method} (\deim) \cite{Chaturantabut2010, Aguado2013}:
\begin{align*}
  \nu \, (\,u\,) \egal \sum_{j\,=\,1}^{K} \, \nu_{\,t}^{\,j} \, (\, x \,) \, \nu_{\,x}^{\,j} \, (\, t \,)\,.
\end{align*}
This decomposition enables us to separate the coefficient into a component depending on the coordinate of the problem. Therefore, Eq.~\eqref{eq:enrichissement_equation} becomes:
\begin{align*}
\begin{split} 
  \frac{\mathrm{d} S}{ \mathrm{d} t} \ R \moins \sum_{j\,=\,1}^{K} \, \nu_{\,t}^{\,j} \ S \ \nu_{\,x}^{\,j} \ \frac{\mathrm{d}^{\,2} R}{ \mathrm{d} x} \egal \sum_{i\,=\,1}^{m} \frac{\mathrm{d} G^{\,i}}{ \mathrm{d} t} \ F^{\,i} \moins \sum_{j\,=\,1}^{K} \, \nu_{\,t}^{\,j} \ G^{\,i} \ \nu_{\,x}^{\,j} \ \frac{\mathrm{d}^{\,2} F^{\,i}}{ \mathrm{d} x^{\,2}} \plus \mathrm{Res}^{\,m+1}\,.
\end{split}
\end{align*}
Functions $R$ and $S$ are then computed using a similar approach as the one described in Section~\ref{sec:PGD}.


\subsection{Case study}

The material investigated is the wood fiber, which properties have been presented in \cite{Rafidiarison2015}. The moisture transport coefficient $\dm$ is assumed as a first-degree polynomial of the relative humidity, while the moisture capacity $\cm$ as a second-degree polynomial:
\begin{align*}
  \cm\,(\,\phi \,) & \egal 120 \, \phi^{\,2} \moins 98 \, \phi \plus 27.02 \,, && \mathsf{kg/m^{\,3}/Pa} \,, \\
  \dm\,(\,\phi \,) & \egal \bigl(\, 5.65 \, \phi \plus 2.33 \bigr) \cdot 10^{\,-11} \,, && \mathsf{s}\,.
\end{align*}

\begin{figure}
  \centering
  \includegraphics[width=0.68\textwidth]{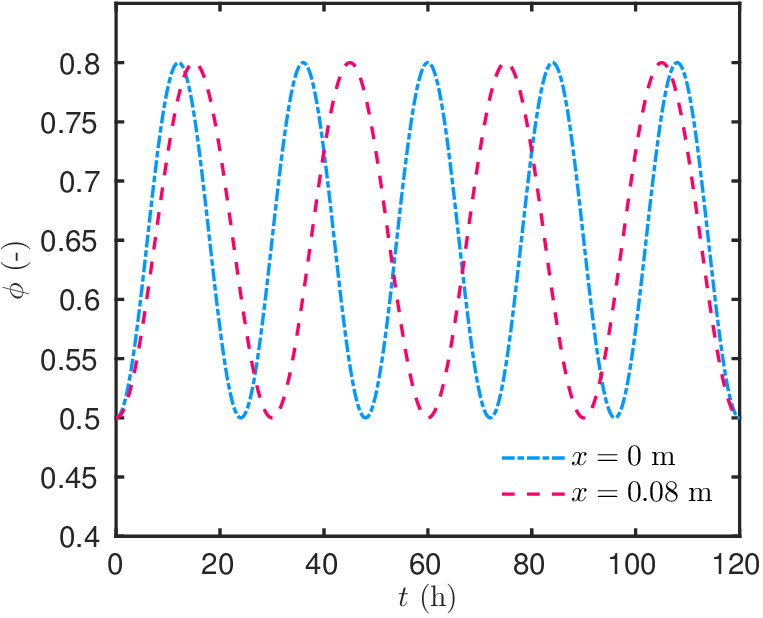}
  \caption{\small\em Time evolution of the boundary condition. }
  \label{fig:caseNL_BC}
\end{figure}

In terms of boundary conditions, a \textsc{Robyn}-type is assumed for both sides of the material, as described in Eq.~\eqref{eq:bc}. The variation of the relative humidity of the ambient air is given in Figure~\ref{fig:caseNL_BC}. Variations were chosen in order to excite the material in the hygroscopic region of the properties. The vapor convective transfer coefficients are set to $\hvL \egal \, 1 \cdot 10^{\,-8} \mathsf{s/m}$ and $\hvR \egal 1.5 \cdot 10^{\,-8} \, \mathsf{s/m}$. As for the previous case, the time simulation is fixed to $5$ days.

\begin{figure}
  \centering
  \subfigure[a][\label{fig:caseNL_profil_phi}]{\includegraphics[width=0.48\textwidth]{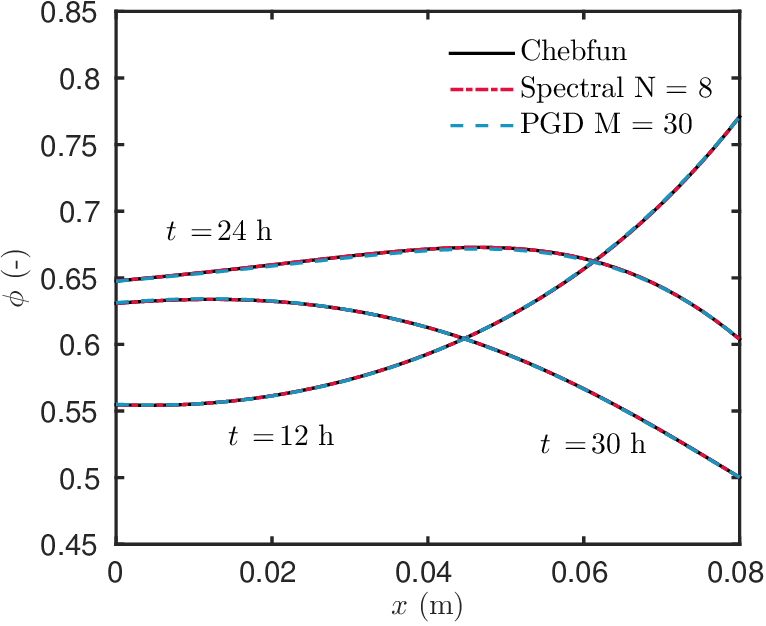}}
  \subfigure[b][\label{fig:caseNL_time_phi}]{\includegraphics[width=0.48\textwidth]{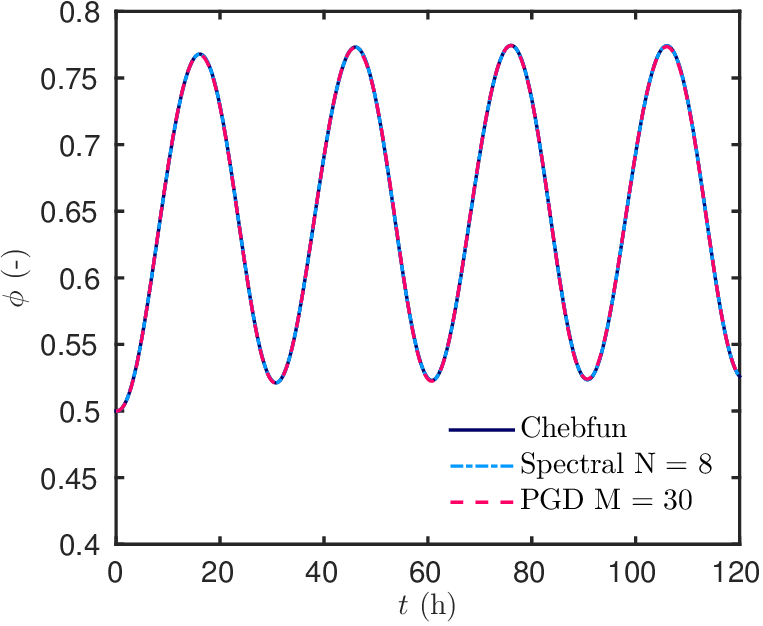}}
  \caption{\small\em (a) Relative humidity profiles in the material and (b) relative humidity evolution at $x \,=\, 0.074 \, \mathsf{m}$.}
\end{figure}

\begin{figure}
  \centering
  \subfigure[a][\label{fig:caseNL_errL2}]{\includegraphics[width=0.48\textwidth]{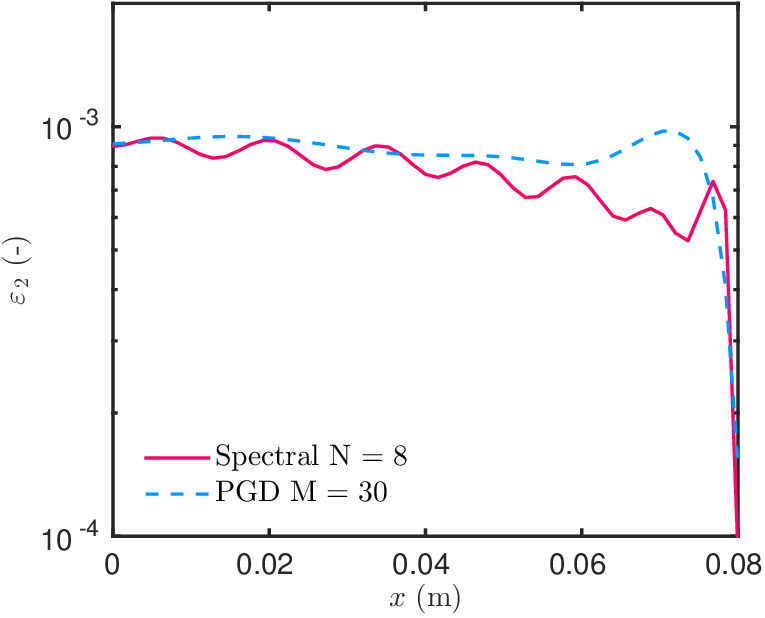}}
  \subfigure[b][\label{fig:caseNL_L2_fModes}]{\includegraphics[width=0.48\textwidth]{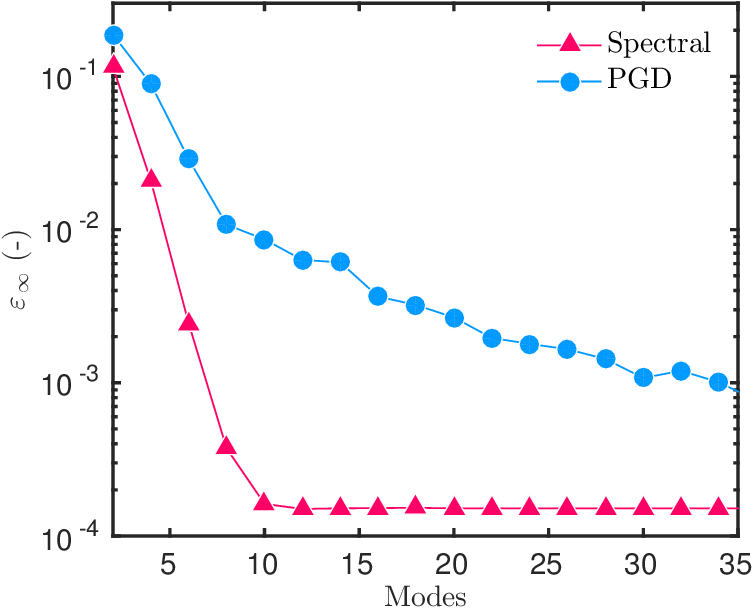}}
  \caption{\small\em (a) Error as a function of $x \, (\mathsf{m})$ and (b) error as a function of the number of modes.}
\end{figure}

Results have been computed using discretization parameters $\Delta \xs \,=\, 10^{\,-2}$ and $\Delta \ts \,=\, 10^{\,-2}$ for both methods. The Spectral--ROM has been built for $N\ =\ 8$ modes while the PGD for $M \ =\ 30$ modes. both results have been compared to a reference solution computed with the \texttt{Matlab} open source toolbox \texttt{Chebfun}. Profiles of relative humidity in the material are shown in Figure~\ref{fig:caseNL_profil_phi}. The time evolution of relative humidity at $x\ =\ 0.074 \, \mathsf{m}$ is given in Figure~\ref{fig:caseNL_time_phi}. A very good agreement is highlighted between the solutions. The physical phenomena are accurately represented. The relative humidity at $x \ =\ 0.074 \, \mathsf{m}$ increases according to the variation of the boundary conditions. The reduced-order models have been built to give the same order of accuracy $\O\,(\,10^{\,-3}\,)$ as illustrated in Figure~\ref{fig:caseNL_errL2}. It can be noted that the PGD needs $M \ =\ 30$ modes to compute the solution of the nonlinear problem, while only $M \ =\ 22$ modes were required in the linear case. The Spectral--ROM only needs one extra mode compared to the linear case. The error with the reference solution is given as a function of the number of modes for both reduced-order models in Figure~\ref{fig:caseNL_L2_fModes}. Again, the Spectral--ROM converges faster to an accurate solution than the PGD approach. A limit is observed in the error of the Spectral--ROM, around $\O\,(\,10^{\,-4}\,)\,$, due to the tolerance of the \texttt{Matlab} solver that was set to this value.

The CPU time of each method has been evaluated for the same order of accuracy of the solution and are reported in Table~\ref{tb:Cpu_time}. For the comparison, the CPU time required with the classical \Eu ~implicit scheme is also indicated. Both methods enable significant computational savings, $95\, \unit{ \% }$ and $99.1 \, \unit{ \% }$ for the PGD and Spectral, respectively. The PGD requires more time than the Spectral--ROM, mainly due to the treatment of the nonlinearity of the problem. It can be noted that using the \deim ~for the treatment of the nonlinearity permits to reduced by more than two the CPU time of the PGD, compared to the \svd. Indeed, in the latter case, at each iteration, the solution has to be composed to evaluate the nonlinear coefficients and then separate them along each coordinate of the problem. The CPU time of each approach is related to the number of operations. For the PGD, it scales with:
\begin{align*}
  & \text{PGD:} && \O\,\Bigl(\, M \cdot \Bigl(\, S \cdot \bigl(\, N_{\,t} \plus N_{\,x} \, \bigr) \plus N_{\,\mathrm{nl}} \, \Bigl) \, \Bigr) \,,
\end{align*}
where $N_{\,\mathrm{nl}}$ represents the number of operations for the treatment of nonlinearities. 
Depending on the method used for the decomposition of the solution, the number of operations scales with:
\begin{align*}
  & \text{\svd:} && \O\,(N_{\,\mathrm{nl}}) \ \simeq \ \Bigl( \, N_{\,x}^{\,2} \cdot N_{\,t} \, \Bigr) \,, \\
  & \text{\deim:} && \O\,(N_{\,\mathrm{nl}}) \ \simeq \ \Bigl(\, K \cdot  (\, N_{\,x} \plus N_{\,t} \,) \, \Bigr) \,,
\end{align*}
where $\O\,(\,K\,) \, \simeq \, 3$ is the order of the decomposition of the coefficients. It can be understood why the CPU time of the PGD using the \deim ~is lower.

For the Spectral--ROM, thanks to the analytical pre-treatment of the solution, there is almost no increase in the number of operations. According to Eq.~\eqref{eq:rearang_diff_nonlin}, the number of operations is only multiplied by two:
\begin{align*}
  & \text{Spectral--ROM:} && \O\,\biggl(\, 2 \ \dfrac{6 \cdot  (N \moins 2)^{\,2} \cdot \tau}{(\mathsf{10^{\,-3}})^{\, 1/5}} \, \biggr) \,.
\end{align*}

For the large original model, using the \Eu ~implicit scheme, the number of operations equals \cite{Gasparin2017}:
\begin{align*}
  & \text{\Eu ~implicit LOM:} && \O\,\Bigl(\, N_{\,\mathrm{nl}} \cdot  N_{\,x} \cdot  N_{\,t} \, \Bigr) \,.
\end{align*}
For this case, the average number of sub-iterations required required for the implicit scheme to treat the nonlinearity was around $ \O\,(N_{\,\mathrm{nl}})\, \simeq\, (\,12\,)$ for a tolerance fixed to $0.01 \cdot \Delta \ts\,$.

\begin{table}
  \centering
  \caption{\small\em Computational cost of the methods for the nonlinear case.} \bigskip
  \label{tb:Cpu_time}
  \setlength{\extrarowheight}{.3em}
  \begin{tabular}{lcc}
  \hline
  Method & CPU time ($\unit{s}$) & CPU time ($\unit{\%}$) \\
  \hline
  \Eu ~implicit & 36   & 100 \\
  PGD using \svd & 5.29  & 15 \\
  PGD using \deim & 1.9  & 5 \\
  Spectral--ROM    & 0.35 & 0.9 \\
  \hline
  \end{tabular}
\end{table}


\section{Conclusion}

Due to moisture-dependent material properties and weather driven boundary conditions, numerical methods are used to compute the solution of moisture transfer problems. Usual approaches based for instance on \textsc{Euler} or \textsc{Crank}--\textsc{Nicolson} schemes require the solution of large systems of equations, which imposes important numerical costs. Model reduction techniques appear then as efficient alternatives, enabling to reduce the model order without deteriorating the representation of the physical phenomena. These methods aim at preserving the computation resources in terms of CPU time and memory. Among the \emph{a priori} model reduction techniques applied for diffusion problems, this paper intended to compare the Spectral and the PGD methods. It provided extensive details in order to provide a numerical benchmark for the community intending to building reduced-order models of diffusion problems in porous material. The case studies provided clear example of the construction and the use of these reduced-order models. Since the dimensionless properties of each case were provided in the Appendix~\ref{sec:append_dimen}, these numerical benchmarks can be used for the whole community of transport in porous media.

The two reduced-order methods assume that the solution is approximated by a finite sum of functions products. The Spectral method fixes a set of basis function for the space domain. Here the \textsc{Chebyshev} polynomials have been chosen. Analytical preliminary treatment of the Spectral solution has been operated to set an ordinary system of equations to compute the temporal coefficients of the solution. On the other hand, the PGD has no assumptions and attempt to compute directly the basis functions by minimizing the equation residual. The comparison was carried out for three cases, commonly found in building physics. The first one deals with linear moisture transfer. The second one aimed at computing a parametric solution, whose model outputs depend not only on the space and time coordinates, but also on the moisture capacity of the material. The last case dealt with a nonlinear transfer problem, with moisture dependent material properties.

Results have demonstrated that both reduced-order models, Spectral and PGD, accurately represent the moisture transfer and both approaches provide an important reduction of the model order. While the order of the large original model scales with several hundred, the one of the ROMs is proportional to a few tens or even less. For the nonlinear case, thanks to this order reduction, the model reduction techniques enable to save more than $95\, \unit{ \% }$ of the CPU time, compared to a large original model based on a \Eu ~implicit scheme. If both methods are efficient, some distinctions between the two ROMs have been highlighted. For the linear and nonlinear cases, the Spectral--ROM has a lower order than the PGD, $\O\,(\,N\,) \, \simeq \, 8$ against $\O\,(\,M\,) \, \simeq \, 20$ to $30\,$, for the same accuracy. Moreover, the error of the Spectral--ROM decreases faster with the number of modes than the PGD. For these reasons, the Spectral--ROM is faster than the PGD. For the nonlinear case, the CPU time of the Spectral--ROM is divided by $5$ compared to the PGD. For the parametric case study, the  two approaches compute the solution by different processes. The Spectral--ROM computes a solution for each numerical value of the material properties within a defined interval. Then, a loop is operated to compute the solution for each value of the material properties. The PGD approach considers directly the material properties as a coordinate of the problem within a defined interval of values. The solution is approximated by a tensorial representation and a basis of functions of each of the three coordinates is computed. Thus, the parametric solution is obtained \emph{at once}. If the PGD ROM needs more modes than the Spectral method, the number of operations to compute the solution is smaller. Moreover, the increase of operations for the PGD is much slower, in this case, thanks to the tensorial representation of the solution. For a parametric solution depending on $50$ values of material properties, the CPU time of the PGD is six times faster.

To conclude the comparison of the two model reduction techniques, results have highlighted that the Spectral approach is more efficient in terms of order reduction, preserving computational resources for linear and nonlinear moisture diffusion problems. For the computation of parametric solutions, the PGD appears to be more efficient. These promising results encourage further investigation for two- or three-dimensional problems including combined heat and moisture transfer phenomena in porous media, where the order of the large original model becomes even higher.


\section*{Acknowledgements}

The authors acknowledge the Brazilian Agencies CAPES of the Ministry of Education and CNPQ of the Ministry of Science, Technology and Innovation, for the financial support.


\section*{Nomenclature}

\begin{tabular*}{0.7\textwidth}{@{\extracolsep{\fill}} |cll| }
\hline
\multicolumn{3}{|c|}{\emph{Latin letters}} \\
$c_{\,m}$ & moisture storage capacity & $[\mathsf{kg/m^3/Pa}]$ \\
$d_{\,m}$ & moisture diffusion & $[\mathsf{s}]$ \\
$g$ & liquid flux & $[\mathsf{kg/m^2/s}]$ \\
$h_{\,v}$ & vapour convective transfer coefficient & $[\mathsf{s/m}]$ \\
$k$ & permeability & $[\mathsf{s}]$ \\
$L$ & length & $[\mathsf{m}]$ \\
$\Pc$ & capillary pressure & $[\mathsf{Pa}]$ \\
$\Ps$ & saturation pressure & $[\mathsf{Pa}]$ \\
$\Pv$ & vapour pressure & $[\mathsf{Pa}]$ \\
$R_v$ & water gas constant & $[\mathsf{J/kg/K}]$\\
$T$ & temperature & $[\mathsf{K}]$ \\
\multicolumn{3}{|c|}{\emph{Greek letters}} \\
$\phi$ & relative humidity & $[-]$ \\
$\rho$ & specific mass & $[\mathsf{kg/m^3}]$ \\ 
\hline
\end{tabular*}


\appendix
\section{Dimensionless values}
\label{sec:append_dimen}

\subsection{Linear case}

Problem~\eqref{eq:moisture_dimensionlesspb_1D} is taken into account with $\glsL \egal \glsR \egal 0 \,$ and a \textsc{Dirichlet} condition on the left side:
\begin{subequations}
  \begin{align}
    \cms \, \pd{u}{\ts} &\egal \pd{}{\xs} \; \left( \, \dms \, \pd{u}{\xs} \, \right) \,,
    & \ts & \ > \ 0\,, \,&  \xs & \ \in \ \big[ \, 0, \, 1 \, \big] \,, \\[3pt]
    u &\egal \uL \,,
    & \ts & \ > \ 0\,, \,&  \xs & \egal 0 \,, \\[3pt]
    \moins \dms \, \pd{u}{\xs} &\egal \BivR \cdot \Bigl( \, u \moins \uR(\,\ts\,) \, \Bigr) \,,
    & \ts & \ > \ 0\,, \,&   \xs & \egal 1 \,, \\[3pt]
    u &\egal 1 \,,
    & ts & \egal 0\,, \,&  \xs & \ \in \ \big[ \, 0, \, 1 \, \big] \,.
  \end{align}
\end{subequations}

The dimensionless properties of the material are $\dms \egal 1$ and $\cms \egal 430\,$. The reference time is $\tref \egal 1$  $\mathsf{h}\,$, thus the final simulation time is fixed to $t^{\,\star} \egal 120\,$. The \textsc{Biot} number is $\BivR \egal 333\,$. The boundary conditions are expressed as:
\begin{align*}
  & \uL  \egal 1 \,, \\ 
  & \uR (\,\ts\,) \egal 1 \plus 1.6 \,\sin^{\,2} \left(\, \frac{2\pi \, \ts}{48}\,\right) \,.
\end{align*}


\subsection{Parametric case}

Problem~\eqref{eq:moisture_dimensionlesspb_1D} is taken into account with $\glsL \egal \glsR \egal 0\,$ and a \textsc{Dirichlet} condition on the left side, the same as in the previous case. The reference time is $\tref \egal 1$  $\mathsf{h}\,$, thus the final simulation time is fixed to $t^{\,\star} \egal 120\,$. The \textsc{Biot} number is $\BivR \egal 100\,$. The boundary conditions are expressed as:
\begin{align*}
  & \uL \egal 1 \,, \\ 
  & \uR (\,\ts\,) \egal 1 \plus 1.6 \,\sin^{\,2} \left(\, \frac{2\pi \, \ts}{48}\,\right) \,.
\end{align*}
The dimensionless properties of the materials are $\dms \egal 1$ and $\cms$ assume the following values:
\begin{center}
  \begin{tabular}{ccccccccccc}
  \hline
  i &1 &2 &3 &4 &5 &6 &7 &8 &9 &10\\
  \hline
  $c_{\,m,i}^{\,\star}$ &833 &576 &441 &357 &300 &258 &227 &203 &183 &166\\
  \hline
  \end{tabular}
\end{center}


\subsection{Nonlinear case}

Problem~\eqref{eq:moisture_dimensionlesspb_1D} is taken into account with $\glsL \egal \glsR \egal 0\,$ and \textsc{Robin} condition on both boundaries. The \textsc{Biot} number are $\BivL \egal 10$ and $\BivR \egal 15\,$. The boundary conditions are expressed as:
\begin{align*}
  & \uL (\,\ts\,)  \egal 1  \plus 0.3 \, \biggl[ \, 1 \moins \cos \left(\, \frac{2\pi \, \ts}{24}\,\right) \, \biggr] \,, \\ 
  & \uR (\,\ts\,) \egal 1 \plus 0.6 \,\sin^{\,2} \left(\, \frac{2\pi \, \ts}{60}\,\right) \,.
\end{align*}

The reference time is $\tref \egal 1$  $\mathsf{h}$, thus the final simulation time is fixed to $t^{\,\star} \egal 120\,$. The dimensionless properties of the materials are:
\begin{align*}
  \dms\, (\,u\,) & \egal \bigl(\, 0.86 \plus 0.25 \, u \, \bigr) \cdot 5 \cdot 10^{\,-3} \,, \\
  \cms\, (\,u\,) & \egal 3.36 \moins 6.11 \, u \plus 3.37 \, u^{\,2} \,.
\end{align*}


\section{Solving the Spectral ODE reduced system}
\label{sec:append_solving_ode}

Consider a more general situation of the reduced-order system~\eqref{eq:system_ode}: 
\begin{equation}\label{eq:system_ode_general}
  \left\{ \begin{array}{rcl}
    \dot{a}\,(\,t\,) & \egal& \A\,(\,\nu\, )\, a \plus \b\, (\,t \,, a\,(\,t\,) \,;\nu \,) \,, \\
    a\,(\,t_{\,0}\,)& \egal& a_{\,0} \,,
  \end{array}\right.
\end{equation}
where $\b\,(\,t \,, a\,(\,t\,) \,;\,\nu \,)$ depends on the solution $a\,(\,t\,)$ via nonlinear boundary conditions, or it contains problem's nonlinearities, if there are some. The dependence on parameters is the most accurate within the chosen Spectral framework. The general analytical solution to problem~\eqref{eq:system_ode_general} can be written as:
\begin{align}\label{eq:solution_ode_general}
  a\,(\,t\,;\nu\,) \egal \mathrm{e}^{\, (\, t \moins t_{\,0} \,) \, \A\, (\,\nu \,) }\, a_{\,0} \plus \int_{t_{\,0}}^{\,t} \, \mathrm{e}^{\, (\,t \moins \tau\,)\, \A\, (\,\nu\,)} \, \b\, \bigl(\, \tau \,, a(\tau)\,; \nu \,\bigr) \, \mathrm{d}\tau\,.
\end{align}
The exponential matrix is defined as the limit:
\begin{align*}
  \mathrm{e}^{\,t\,\A} \egal \lim_{n\, \rightarrow \, \infty} \biggl(\,\mathrm{Id} \plus \dfrac{1}{n} \, t \, \A \, \biggr)^{\,n} \,,
\end{align*}
in which $\mathrm{Id} \in \Mat_{(n-2)\times (n-2)}(\R )\,$ is the identity matrix. However, this method is not the best way to compute the exponential matrix. In some particular cases, the solution of Eq.~\eqref{eq:system_ode_general} can be simplified and thus better exploited \cite{Moler2003}.


\bigskip
\paragraph{Case I:} 

If we have homogeneous boundary conditions, problem \eqref{eq:system_ode_general} becomes:
\begin{align*}
  \left\{ \begin{array}{rcl}
    \dot{a}& \egal& \A\, (\,\nu\,) \, a \,,\\
    a\,(\, t_{\,0} \,)& \egal& a_{\,0} \,,
  \end{array}\right. 
\end{align*}
and it can be analytically solved as:
\begin{align*}
  a\,(\,t \,; \nu\,) \egal \mathrm{e}^{\,(\, t \moins t_{\,0} \,)\, \A\,(\,\nu\,)} \, a_{\,0}\,.
\end{align*}

Using modern methods, the exponential matrix can be computed using $\sim 48\,n^{\,3}$ \emph{floating point operations per second} (FLOPS) \cite{Al-Mohy2010}. As an information to the reader, the previous result was $\sim 538\,n^{\,3}$ FLOPS \cite{Kenney1998}.

However, one can notice that we do not really need to build the exponential matrix, but we want to compute its \textit{action} on the initial state vector $a_{\,0}$. Nowadays, it can be directly done, without forming $\mathrm{e}^{\,(\, t \moins t_{\,0}\,)\,\A\, (\, \nu\, )}\,$, explicitly to a prescribed accuracy that can be set significantly lower than the standard machine precision $\sim 10^{\,-16}$ \cite{Al-Mohy2011}. If computing Eq.~\eqref{eq:system_ode_general} by a \texttt{Matlab} solver, for example \texttt{ODE45}, the standard tolerance is of order of $\sim 10^{-6}\,$.


\bigskip
\paragraph{Case II:}

If we have inhomogeneous boundary conditions constant in time, the problem from Eq.~\eqref{eq:system_ode_general} becomes:
\begin{align*}
  \left\{ \begin{array}{rcl}
    \dot{a}& \egal& \A\, (\, \nu\, )\, a \plus \b\, (\, \nu\, ) \,, \\
    a\, (\, t_{\,0}\, )& \egal& a_{\,0} \,,
  \end{array}\right. 
\end{align*}
which can also be analytically solved: 
\begin{align*}
  a\, (\, t\,; \nu\,) \egal \mathrm{e}^{\,( \, t \moins t_{\,0} \,) \, \A\, (\,\nu\, )} \, a_{\,0}  \plus (\, t \moins t_{\,0} \,) \, \frac{\mathrm{e}^{\,( \, t \moins t_{\,0} \,) \, \A\, (\, \nu\, )} \moins \mathrm{Id} }{(\, t \moins t_{\,0} \,) \, \A\, (\,\nu\, )} \, \b\, (\, \nu\, ) \,.
\end{align*}


\bigskip
\paragraph{Case III:} 

If we have inhomogeneous boundary conditions are linear in time, problem Eq.~\eqref{eq:system_ode_general} becomes:
\begin{align*}
  \left\{ \begin{array}{rcl}
  \dot{a} & \egal & \A\, (\, \nu\, ) \, a \plus \b\, (\, \nu\, )\cdot t  \,, \\
  a\, (\, t_{\,0}\, )& \egal& a_{\,0} \,,
  \end{array}\right.
\end{align*}
the solution is given by: 
\begin{align*}
  a\, (\, t\,; \nu\,) \egal \mathrm{e}^{\,(\, t \moins t_{\,0} \,) \, \A\, (\, \nu\, )} \, a_{\,0}  \plus ( \, t \moins t_{\,0} \,)^{\,2} \ \frac{\mathrm{e}^{\,( \, t \moins t_{\,0} \,) \, \A\, (\, \nu\, )} \moins \mathrm{Id} }{( \, t \moins t_{\,0} \,) \, \A\, (\, \nu\, )} \ \b\, (\, \nu\, ) \,.
\end{align*}


\bigskip
\paragraph{Case IV:}

Boundary condition is polynomial in time:
\begin{align*}
  \left\{ \begin{array}{rclr}
  \dot{a}& \egal& \A\, (\, \nu\, ) \, a \plus \b\, (\, \nu\, ) \cdot t^{\, m-1}\,, & m \ \geqslant \ 3 \,, \\
  a\, (\, t_{\,0}\, )& \egal& a_{\,0} \,.
  \end{array}\right. 
\end{align*}
Then, the solution is given by the following analytical formula: 
\begin{align*}
  a\, (\, t\,; \nu\,) \egal \mathrm{e}^{\,( \, t \moins t_{\,0} \,) \, \A\, (\, \nu\, )} \, a_{\,0}  \plus ( \, t \moins t_{\,0} \,)^{\,m} \ \varphi_{\, m} \Bigl(\, ( \, t \moins t_{\,0} \,) \, \A\, (\, \nu\, )\, \Bigr) \, \b\, (\, \nu\, ) \,.
\end{align*}

Above we introduced the so-called matrix $\varphi-$functions:
\begin{align*}
  & \varphi_{\,m}\, (\, z \,) \egal \dfrac{\varphi_{\,m}\, (\, z\,) \moins \dfrac{1}{m\,!}}{z}\,, & m \ \geqslant \ 0 \quad \text{with} \quad \varphi_{\,0}\, (\,z\,)\ \equiv\ \mathrm{e}^{\,z} \,.
\end{align*}

A few first functions are given below explicitly:
\begin{align*}
  \varphi_{\, 0}\, (\, z \,) &\egal \mathrm{e}^{\,z} \egal 1 \plus z \plus \dfrac{1}{2}\, z^{\,2} \plus \dfrac{1}{3\,!}\, z^{\,3} \plus \ldots \\
  \varphi_{\, 1}\, (\, z \,) &\egal \dfrac{\mathrm{e}^{\,z}-1}{z} \egal 1 \plus \dfrac{1}{2}\, z \plus \dfrac{1}{3\,!}\, z^{\,2} \plus \dfrac{1}{4\,!}\, z^{\,3}   \plus    \ldots \\
  \varphi_{\, 2}\, (\, z \,) &\egal \dfrac{\mathrm{e}^{\,z}-1-z}{z^{\,2}} \egal \dfrac{1}{2} \plus \dfrac{1}{3\,!}\, z \plus \dfrac{1}{4\,!}\, z^{\,2} \plus \dfrac{1}{5\,!}\, z^{\,3} \plus \ldots \\
  \varphi_{\, 3}\, (\, z \,) &\egal \dfrac{\mathrm{e}^{\,z}-1-z-\frac{1}{2}}{z^{\,3}} \egal \dfrac{1}{3\,!} \plus \dfrac{1}{4\,!}\, z \plus \dfrac{1}{5\,!}\, z^{\,2} \plus \dfrac{1}{6\,!}\, z^{\,3} \plus \ldots 
\end{align*}

The general power series representation of $\varphi-$functions is
\begin{align*}
  & \varphi_{\, m}\, (\, z \,)\ \equiv\ \sum_{k\; =\; 0}^{\,\infty}\, \dfrac{z^{\, k}}{(\, m \plus k\, )\,!}\,.
\end{align*}
The exponential definitions of $\varphi_{\, m}\, (\, z \,)$ should not be used for practical simulations, because of severe cancellation errors for $z\, \ll\, 1\,$. Efficient methods for computation of $\varphi-$functions have been developed based on \textsc{Pad\'e}-type expansions, to give an example, \texttt{Matlab}'s function \texttt{expm()} is based on such approximations \cite{Cox2002}.


\bigskip
\paragraph{Case V:}

For a general case of linear boundary conditions, the solution of problem Eq.~\eqref{eq:system_ode_general} is:
\begin{align*}
  a\, (\, t\,; \nu\,) \egal \mathrm{e}^{\, (\, t \moins t_{\,0}\, )\, \A\, (\,\nu\, )}\, a_{\,0} \plus &\underbrace{\int_{t_{\,0}}^{\,t}\, \mathrm{e}^{\, (\,t \moins \tau\, )\, \A\, (\,\nu\, )} \, \b\, (\, \tau \,; \nu\, ) \, \mathrm{d}\tau}_{(I)}\,.
\end{align*}
To exploit the last formula, one might employ a quadrature formula to discretize the integral (I):
\begin{align*}
  a\, (\, t\,; \nu\,) \egal \mathrm{e}^{\, (\, t \moins t_{\,0}\, )\, \A\, (\,\nu\, )}\,  a_{\,0} \plus \Delta t \, \sum^{m}_{j\; =\;1}\, \mathrm{e}^{\, (\, t \moins t_{\,0}\, )\, \A\, (\,\nu\, )} \cdot \b\, (\, \tau_{\, j} \,; \nu\, )\,, & & \Delta t\ \eqdef\ \frac{t \moins t_{\,0}}{m}\,,
\end{align*} 
where we employed rectangle formula for simplicity. We note that the sequence $\{\mathrm{e}^{\, \Delta t \, \A\, (\,\nu\, )} \}_{\,j\,=\,1}^{\,m}$ can be entirely computed in an efficient manner \cite{Al-Mohy2011}.


\bigskip
\paragraph{Case VI:}

Considering a general nonlinear case of boundary conditions from problem Eq.~\eqref{eq:system_ode_general} and the general solution Eq.~\eqref{eq:solution_ode_general}. To exploit a better solution, we can develop the function $\tau\, \mapsto\, \b\, (\,\tau\,, a(\tau)\,; \nu\,)$ in \textsc{Taylor} expansion series and integrate it exactly:
\begin{align*}
  a\, (\, t\,; \nu\,) \egal \mathrm{e}^{\, (\, t \moins t_{\,0}\, )\, \A\, (\,\nu\, )}\, a_{\,0} \plus \sum^{\infty}_{k\; =\;1} (\, t \moins t_{\,0}\, )^{\,k} \, \varphi_{\, k} \Bigl(\, (\, t\moins t_{\,0}\, )\, \A\, (\,\nu\, )\, \Bigr) \, a_{\,k} \,, 
\end{align*}
where,
\begin{align*}
  a_{\,k}\ \eqdef\ \left. \,\dfrac{\mathrm{d}^{\, k\, -\, 1}}{\mathrm{d}t^{\, k \, -\, 1}}\ \b \, \Bigl(\, t\,, a(\,t\,)\,;\nu \, \Bigr) \, \right|_{\, t \egal t_{\,0}} \,.
\end{align*}

Finally, the series solution can be exploited by truncating it at some finite order:
\begin{align*}
  a\, (\, t\,; \nu\,) \egal \mathrm{e}^{\, (\, t \moins t_{\,0}\, )\, \A\, (\,\nu\, )}\, a_{\,0} \plus \sum_{k\; =\;1}^{K} (\, t \moins t_{\,0}\, )^{\,k} \, \varphi_{\,k} \Bigl(\, (\, t \moins t_{\,0}\, )\, \A\, (\,\nu\, )\, \Bigr) \, a_{\,k} \,.
\end{align*}


\bigskip
\addcontentsline{toc}{section}{References}
\bibliographystyle{abbrv}
\bibliography{biblio}

\begin{thebibliography}{10}

\bibitem{Aguado2013}
J.~V. Aguado, F.~Chinesta, A.~Leygue, E.~{Cueto Prendes}, and A.~Huerta.
\newblock {Deim-based PGD for parametric nonlinear model order reduction}.
\newblock In {\em International Conference on Adaptive Modeling and
  Simulation}, pages 29--34. Centre Internacional de M{\`{e}}todes
  Num{\`{e}}rics en Enginyeria (CIMNE), 2013.

\bibitem{AitOumeziane2014}
Y.~{A{\"{i}}t Oumeziane}, M.~Bart, S.~Moissette, and C.~Lanos.
\newblock {Hysteretic Behaviour and Moisture Buffering of Hemp Concrete}.
\newblock {\em Transport in Porous Media}, 103(3):515--533, jul 2014.

\bibitem{Al-Mohy2010}
A.~H. Al-Mohy and N.~J. Higham.
\newblock {A New Scaling and Squaring Algorithm for the Matrix Exponential}.
\newblock {\em SIAM J. Matrix Anal. Appl.}, 31(3):970--989, jan 2010.

\bibitem{Al-Mohy2011}
A.~H. Al-Mohy and N.~J. Higham.
\newblock {Computing the Action of the Matrix Exponential, with an Application
  to Exponential Integrators}.
\newblock {\em SIAM J. Sci. Comput.}, 33(2):488--511, jan 2011.

\bibitem{Al-Sanea2011}
S.~A. Al-Sanea and M.~F. Zedan.
\newblock {Improving thermal performance of building walls by optimizing
  insulation layer distribution and thickness for same thermal mass}.
\newblock {\em Applied Energy}, 88(9):3113--3124, sep 2011.

\bibitem{Al-Sanea2005}
S.~A. Al-Sanea, M.~F. Zedan, and S.~A. Al-Ajlan.
\newblock {Effect of electricity tariff on the optimum insulation-thickness in
  building walls as determined by a dynamic heat-transfer model}.
\newblock {\em Applied Energy}, 82(4):313--330, dec 2005.

\bibitem{Ammar2008}
A.~Ammar and F.~Chinesta.
\newblock {Circumventing Curse of Dimensionality in the Solution of Highly
  Multidimensional Models Encountered in Quantum Mechanics Using Meshfree
  Finite Sums Decomposition}.
\newblock In {\em Meshfree Methods for Partial Differential Equations IV},
  pages 1--17. Springer Berlin Heidelberg, Berlin, Heidelberg, 2008.

\bibitem{Ammar2007}
A.~Ammar, B.~Mokdad, F.~Chinesta, and R.~Keunings.
\newblock {A new family of solvers for some classes of multidimensional partial
  differential equations encountered in kinetic theory modelling of complex
  fluids}.
\newblock {\em J. Non-Newtonian Fluid Mech.}, 144(2-3):98--121, jul 2007.

\bibitem{Ammar2010}
A.~Ammar, M.~Normandin, F.~Daim, D.~Gonzalez, E.~Cueto, and F.~Chinesta.
\newblock {Non incremental strategies based on separated representations:
  applications in computational rheology}.
\newblock {\em Comm. Math. Sci.}, 8(3):671--695, 2010.

\bibitem{Aste2009}
N.~Aste, A.~Angelotti, and M.~Buzzetti.
\newblock {The influence of the external walls thermal inertia on the energy
  performance of well insulated buildings}.
\newblock {\em Energy and Buildings}, 41(11):1181--1187, nov 2009.

\bibitem{Axaopoulos2014}
I.~Axaopoulos, P.~Axaopoulos, and J.~Gelegenis.
\newblock {Optimum insulation thickness for external walls on different
  orientations considering the speed and direction of the wind}.
\newblock {\em Applied Energy}, 117:167--175, mar 2014.

\bibitem{Bednar2005}
T.~Bednar and C.-E. Hagentoft.
\newblock {Analytical solution for moisture buffering effect Validation
  exercises for simulation tools}.
\newblock In {\em 7th Nordic Symposium on Building Physics}, Reykjavik,
  Iceland, 2005.

\bibitem{Berger2015}
J.~Berger, M.~Chhay, S.~Guernouti, and M.~Woloszyn.
\newblock {Proper generalized decomposition for solving coupled heat and
  moisture transfer}.
\newblock {\em Journal of Building Performance Simulation}, 8(5):295--311, sep
  2015.

\bibitem{Berger2017d}
J.~Berger and N.~Mendes.
\newblock {An innovative method for the design of high energy performance
  building envelopes}.
\newblock {\em Applied Energy}, 190:266--277, mar 2017.

\bibitem{Berger2016c}
J.~Berger, N.~Mendes, S.~Guernouti, M.~Woloszyn, and F.~Chinesta.
\newblock {Review of Reduced Order Models for Heat and Moisture Transfer in
  Building Physics with Emphasis in PGD Approaches}.
\newblock {\em Archives of Computational Methods in Engineering}, pages 1--13,
  jul 2016.

\bibitem{Bognet2012}
B.~Bognet, F.~Bordeu, F.~Chinesta, A.~Leygue, and A.~Poitou.
\newblock {Advanced simulation of models defined in plate geometries: 3D
  solutions with 2D computational complexity}.
\newblock {\em Comp. Meth. Appl. Mech. Eng.}, 201-204:1--12, jan 2012.

\bibitem{Bond2013}
D.~E.~M. Bond, W.~W. Clark, and M.~Kimber.
\newblock {Configuring wall layers for improved insulation performance}.
\newblock {\em Applied Energy}, 112:235--245, dec 2013.

\bibitem{Boyd2000}
J.~P. Boyd.
\newblock {\em {Chebyshev and Fourier Spectral Methods}}.
\newblock New York, 2nd edition, 2000.

\bibitem{Canuto2006}
C.~Canuto, M.~Y. Hussaini, A.~Quarteroni, and T.~A. Zang.
\newblock {\em {Spectral Methods Fundamentals in Single Domains}}.
\newblock Scientific Computation. Springer-Verlag Berlin Heidelberg, 2006.

\bibitem{Chaturantabut2010}
S.~Chaturantabut and D.~C. Sorensen.
\newblock {Nonlinear Model Reduction via Discrete Empirical Interpolation}.
\newblock {\em SIAM J. Sci. Comput.}, 32(5):2737--2764, jan 2010.

\bibitem{Chen2015b}
S.-S. Chen, B.-W. Li, and Y.-S. Sun.
\newblock {Chebyshev collocation spectral method for solving radiative transfer
  with the modified discrete ordinates formulations}.
\newblock {\em Int. J. Heat Mass Transfer}, 88:388--397, sep 2015.

\bibitem{Chen2016}
Y.-Y. Chen, B.-W. Li, and J.-K. Zhang.
\newblock {Spectral collocation method for natural convection in a square
  porous cavity with local thermal equilibrium and non-equilibrium models}.
\newblock {\em Int. J. Heat Mass Transfer}, 96:84--96, may 2016.

\bibitem{Chinesta2011a}
F.~Chinesta, A.~Ammar, A.~Leygue, and R.~Keunings.
\newblock {An overview of the proper generalized decomposition with
  applications in computational rheology}.
\newblock {\em J. Non-Newtonian Fluid Mech.}, 166(11):578--592, jun 2011.

\bibitem{Chinesta2013a}
F.~Chinesta, R.~Keunings, and A.~Leygue.
\newblock {\em {The Proper Generalized Decomposition for Advanced Numerical
  Simulations: A Primer}}.
\newblock Springer International Publishing, New York, 2013.

\bibitem{Chinesta2011}
F.~Chinesta, P.~Ladev{\`{e}}ze, and E.~Cueto.
\newblock {A Short Review on Model Order Reduction Based on Proper Generalized
  Decomposition}.
\newblock {\em Archives of Computational Methods in Engineering},
  18(4):395--404, nov 2011.

\bibitem{Chinesta2013}
F.~Chinesta, A.~Leygue, F.~Bordeu, J.~V. Aguado, E.~Cueto, D.~Gonzalez,
  I.~Alfaro, A.~Ammar, and A.~Huerta.
\newblock {PGD-Based Computational Vademecum for Efficient Design, Optimization
  and Control}.
\newblock {\em Archives of Computational Methods in Engineering}, 20(1):31--59,
  mar 2013.

\bibitem{Cox2002}
S.~M. Cox and P.~C. Matthews.
\newblock {Exponential Time Differencing for Stiff Systems}.
\newblock {\em J. Comp. Phys.}, 176(2):430--455, mar 2002.

\bibitem{Dalgliesh2005}
A.~Dalgliesh, S.~Cornick, W.~Maref, and P.~Mukhopadhyaya.
\newblock {Hygrothermal Performance of Building Envelopes: Uses for 2D and 1D
  simulation}.
\newblock In {\em 10th Conference on Building Science and Technology}, Ottawa,
  Canada, 2005. NRC Publication Archive.

\bibitem{DosSantos2006}
G.~H. {Dos Santos} and N.~Mendes.
\newblock {Simultaneous heat and moisture transfer in soils combined with
  building simulation}.
\newblock {\em Energy and Buildings}, 38(4):303--314, 2006.

\bibitem{Driscoll2014}
T.~A. Driscoll, N.~Hale, and L.~N. Trefethen.
\newblock {Chebfun Guide}.
\newblock {\em Pafnuty Publications}, Oxford, 2014.

\bibitem{Dumon2011}
A.~Dumon, C.~Allery, and A.~Ammar.
\newblock {Proper general decomposition (PGD) for the resolution of
  Navier-Stokes equations}.
\newblock {\em J. Comp. Phys.}, 230(4):1387--1407, 2011.

\bibitem{Gasparin2018}
S.~Gasparin, J.~Berger, D.~Dutykh, and N.~Mendes.
\newblock {Solving nonlinear diffusive problems in buildings by means of a
  Spectral reduced-order model}.
\newblock {\em Journal of Building Performance Simulation}, 2018.

\bibitem{Gasparin2017}
S.~Gasparin, J.~Berger, D.~Dutykh, and N.~Mendes.
\newblock {Stable explicit schemes for simulation of nonlinear moisture
  transfer in porous materials}.
\newblock {\em J. Building Perf. Simul.}, 11(2):129--144, 2018.

\bibitem{Gautschi2004}
W.~Gautschi.
\newblock {\em {Orthogonal Polynomials: Computation and Approximation}}.
\newblock Oxford University Press, Oxford, UK, 2004.

\bibitem{Golub1996}
G.~Golub and C.~{Van Loan}.
\newblock {\em {Matrix Computations}}.
\newblock J. Hopkins University Press, 3rd ed. edition, 1996.

\bibitem{Guo2012}
W.~Guo, G.~Labrosse, and R.~Narayanan.
\newblock {\em {The Application of the Chebyshev-Spectral Method in Transport
  Phenomena}}, volume~68 of {\em Lecture Notes in Applied and Computational
  Mechanics}.
\newblock Springer Berlin Heidelberg, Berlin, Heidelberg, 2012.

\bibitem{Ibrahim2015}
M.~Ibrahim, P.~H. Biwole, P.~Achard, E.~Wurtz, and G.~Ansart.
\newblock {Building envelope with a new aerogel-based insulating rendering:
  Experimental and numerical study, cost analysis, and thickness optimization}.
\newblock {\em Applied Energy}, 159:490--501, dec 2015.

\bibitem{Janssen2014}
H.~Janssen.
\newblock {Simulation efficiency and accuracy of different moisture transfer
  potentials}.
\newblock {\em J. Building Perf. Simul.}, 7(5):379--389, sep 2014.

\bibitem{Janssen2007}
H.~Janssen, B.~Blocken, and J.~Carmeliet.
\newblock {Conservative modelling of the moisture and heat transfer in building
  components under atmospheric excitation}.
\newblock {\em Int. J. Heat Mass Transfer}, 50(5-6):1128--1140, mar 2007.

\bibitem{Kenney1998}
C.~S. Kenney and A.~J. Laub.
\newblock {A Schur-Fr{\'{e}}chet Algorithm for Computing the Logarithm and
  Exponential of a Matrix}.
\newblock {\em SIAM J. Matrix Anal. Appl.}, 19(3):640--663, jul 1998.

\bibitem{Labat2016}
M.~Labat, C.~Magniont, N.~Oudhof, and J.-E. Aubert.
\newblock {From the experimental characterization of the hygrothermal
  properties of straw-clay mixtures to the numerical assessment of their
  buffering potential}.
\newblock {\em Building and Environment}, 97:69--81, feb 2016.

\bibitem{Ladeveze1985}
P.~Ladev{\`{e}}ze.
\newblock {Sur une famille d'algorithmes en m{\'{e}}canique des structures}.
\newblock {\em Comptes-rendus des s{\'{e}}ances de l'Acad{\'{e}}mie des
  sciences. S{\'{e}}rie 2, M{\'{e}}canique-physique, chimie, sciences de
  l'univers, sciences de la terre}, 300(2):41--44, 1985.

\bibitem{Lamari2012}
H.~Lamari, A.~Ammar, A.~Leygue, and F.~Chinesta.
\newblock {On the solution of the multidimensional Langer's equation using the
  proper generalized decomposition method for modeling phase transitions}.
\newblock {\em Modelling and Simulation in Materials Science and Engineering},
  20(1):7--15, jan 2012.

\bibitem{Li2008a}
B.-W. Li, Y.-S. Sun, and Y.~Yu.
\newblock {Iterative and direct Chebyshev collocation spectral methods for
  one-dimensional radiative heat transfer}.
\newblock {\em Int. J. Heat Mass Transfer}, 51(25-26):5887--5894, dec 2008.

\bibitem{Luikov1966}
A.~V. Luikov.
\newblock {\em {Heat and mass transfer in capillary-porous bodies}}.
\newblock Pergamon Press, New York, 1966.

\bibitem{Ma2014}
J.~Ma, B.-W. Li, and J.~R. Howell.
\newblock {Thermal radiation heat transfer in one- and two-dimensional
  enclosures using the spectral collocation method with full spectrum
  k-distribution model}.
\newblock {\em International Journal of Heat and Mass Transfer}, 71:35--43, apr
  2014.

\bibitem{Mendes2017}
N.~Mendes, M.~Chhay, J.~Berger, and D.~Dutykh.
\newblock {\em {Numerical methods for diffusion phenomena in building
  physics}}.
\newblock PUCPRess, Curitiba, Parana, 2017.

\bibitem{Mendes2005}
N.~Mendes and P.~C. Philippi.
\newblock {A method for predicting heat and moisture transfer through
  multilayered walls based on temperature and moisture content gradients}.
\newblock {\em Int. J. Heat Mass Transfer}, 48(1):37--51, 2005.

\bibitem{Moler2003}
C.~Moler and C.~{Van Loan}.
\newblock {Nineteen Dubious Ways to Compute the Exponential of a Matrix,
  Twenty-Five Years Later}.
\newblock {\em SIAM Review}, 45(1):3--49, jan 2003.

\bibitem{Motsa2015}
S.~Motsa.
\newblock {On the New Bivariate Local Linearisation Method for Solving Coupled
  Partial Differential Equations in Some Applications of Unsteady Fluid Flows
  with Heat and Mass Transfer}.
\newblock In M.~Solecki, editor, {\em Mass Transfer - Advancement in Process
  Modelling}. InTech, Rijeka, oct 2015.

\bibitem{Neron2010}
D.~N{\'{e}}ron and P.~Ladev{\`{e}}ze.
\newblock {Proper Generalized Decomposition for Multiscale and Multiphysics
  Problems}.
\newblock {\em Arch. Comp. Meth. Eng.}, 17(4):351--372, dec 2010.

\bibitem{Niroomandi2012}
S.~Niroomandi, I.~Alfaro, E.~Cueto, and F.~Chinesta.
\newblock {Accounting for large deformations in real-time simulations of soft
  tissues based on reduced-order models}.
\newblock {\em Computer Methods and Programs in Biomedicine}, 105(1):1--12, jan
  2012.

\bibitem{Nouy2007}
A.~Nouy.
\newblock {A generalized spectral decomposition technique to solve a class of
  linear stochastic partial differential equations}.
\newblock {\em Comp. Meth. Appl. Mech. Eng.}, 196(45-48):4521--4537, sep 2007.

\bibitem{Ozel2011}
M.~Ozel.
\newblock {Effect of wall orientation on the optimum insulation thickness by
  using a dynamic method}.
\newblock {\em Applied Energy}, 88(7):2429--2435, jul 2011.

\bibitem{Ozisik1993}
M.~N. Ozisik.
\newblock {\em {Heat conduction}}.
\newblock Wiley-Interscience, New York, 2 edition, 1993.

\bibitem{Pasban2017}
A.~Pasban, H.~Sadrnia, M.~Mohebbi, and S.~A. Shahidi.
\newblock {Spectral method for simulating 3D heat and mass transfer during
  drying of apple slices}.
\newblock {\em Journal of Food Engineering}, 212(Supplement C):201--212, nov
  2017.

\bibitem{Peyret2002}
R.~Peyret.
\newblock {\em {Spectral methods for incompressible viscous flow}}.
\newblock Springer-Verlag, New York, 2002.

\bibitem{Philip1957}
J.~R. Philip and D.~A. {De Vries}.
\newblock {Moisture movement in porous materials under temperature gradients}.
\newblock {\em Transactions, American Geophysical Union}, 38(2):222--232, 1957.

\bibitem{Pruliere2010}
E.~Pruliere, F.~Chinesta, and A.~Ammar.
\newblock {On the deterministic solution of multidimensional parametric models
  using the Proper Generalized Decomposition}.
\newblock {\em Math. Comp. Simul.}, 81(4):791--810, dec 2010.

\bibitem{Rafidiarison2015}
H.~Rafidiarison, R.~R{\'{e}}mond, and E.~Mougel.
\newblock {Dataset for validating 1-D heat and mass transfer models within
  building walls with hygroscopic materials}.
\newblock {\em Building and Environment}, 89:356--368, jul 2015.

\bibitem{RamReddy2015}
C.~RamReddy, P.~A. {Lakshmi Narayana}, and S.~S. Motsa.
\newblock {A spectral relaxation method for linear and non-linear
  stratification effects on mixed convection in a porous medium}.
\newblock {\em Appl. Math. Comput.}, 268:991--1000, oct 2015.

\bibitem{Rode2006}
C.~Rode and R.~H. Peuhkur.
\newblock {The Concept of Moisture Buffer Value of Building Materials and its
  Application in Building Design}.
\newblock In {\em Healthy Buildings 2006}, pages 57--62. 2006.

\bibitem{Roels2003}
S.~Roels, J.~Carmeliet, and H.~Hens.
\newblock {Modelling Unsaturated Moisture Transport in Heterogeneous
  Limestone}.
\newblock {\em Transport in Porous Media}, 52(3):333--350, 2003.

\bibitem{Shampine1997}
L.~F. Shampine and M.~W. Reichelt.
\newblock {The MATLAB ODE Suite}.
\newblock {\em SIAM J. Sci. Comput.}, 18:1--22, 1997.

\bibitem{Trefethen1996}
L.~N. Trefethen.
\newblock {\em {Finite Difference and Spectral Methods for Ordinary and Partial
  Differential Equations}}.
\newblock Unpublished, Ithaca, NY, USA, 1996.

\bibitem{Waldrop2016}
M.~M. Waldrop.
\newblock {The chips are down for Moore's law}.
\newblock {\em Nature}, 530(7589):144--147, feb 2016.

\bibitem{Wang2016}
C.~Wang, Z.~Qiu, and Y.~Yang.
\newblock {Collocation methods for uncertain heat convection-diffusion problem
  with interval input parameters}.
\newblock {\em Int. J. Therm. Sci.}, 107:230--236, sep 2016.

\bibitem{Woloszyn2008}
M.~Woloszyn and C.~Rode.
\newblock {Tools for performance simulation of heat, air and moisture
  conditions of whole buildings}.
\newblock {\em Building Simulation}, 1(1):5--24, mar 2008.

\bibitem{Yuan2016}
J.~Yuan, C.~Farnham, K.~Emura, and M.~A. Alam.
\newblock {Proposal for optimum combination of reflectivity and insulation
  thickness of building exterior walls for annual thermal load in Japan}.
\newblock {\em Building and Environment}, 103:228--237, jul 2016.

\end{thebibliography}
\bigskip\bigskip


\end{document}